\documentclass[fleqn,usenatbib]{mnras}

\pdfoutput=1

\usepackage{newtxtext,newtxmath}

\usepackage[T1]{fontenc}

\usepackage{graphicx}	
\usepackage{amsmath}	

\usepackage{xcolor}

\usepackage{hyperref}
\title[miniJPAS: mocks for quasar identification]{The miniJPAS survey quasar selection I: Mock catalogues for classification}

\author[Queiroz et al.]{
Carolina Queiroz$^{1,2}$\thanks{Email: c.queirozabs@gmail.com},
L. Raul Abramo$^{2}$,
Nat\'alia V. N. Rodrigues$^{2}$,
Ignasi P\'erez-R\`afols$^{3,4}$
\newauthor
Gin\'es Mart\'inez-Solaeche$^{5}$,
Antonio Hern\'an-Caballero$^{6}$,
Carlos Hern\'andez-Monteagudo$^{7,8}$,
\newauthor
Alejandro Lumbreras-Calle$^{6}$,
Matthew M. Pieri$^{4}$,
Sean S. Morrison$^{4,9}$,
Silvia Bonoli$^{10,11}$,
\newauthor
Jon\'as Chaves-Montero$^{10}$,
Ana L. Chies-Santos$^{1,12}$,
L. A. D\'iaz-Garc\'ia$^{5}$,
\newauthor
Alberto Fernandez-Soto$^{13,14}$,
Rosa M. Gonz\'alez Delgado$^{5}$,
Jailson Alcaniz$^{15}$,
\newauthor
Narciso Ben\'itez$^{5}$,
A. Javier Cenarro$^{16}$,
Tamara Civera$^{6}$,
Renato A. Dupke$^{15,17,18}$,
\newauthor
Alessandro Ederoclite$^{6}$,
Carlos L\'opez-Sanjuan$^{16}$,
Antonio Mar\'in-Franch$^{16}$,
\newauthor
Claudia Mendes de Oliveira$^{19}$,
Mariano Moles$^{5,6}$,
David Muniesa$^{16}$,
Laerte Sodr\'e Jr.$^{19}$,
\newauthor
Keith Taylor$^{20}$,
Jes\'us Varela$^{16}$
and H\'ector V\'azquez Rami\'o$^{16}$
\\
$^{1}$ Departamento de Astronomia, Instituto de F\'isica, Universidade Federal do Rio Grande do Sul (UFRGS), Av. Bento Gon\c{c}alves, 9500, Porto Alegre, RS, Brazil\\
$^{2}$ Departamento de F\'isica Matem\'atica, Instituto de F\'{\i}sica, Universidade de S\~ao Paulo, Rua do Mat\~ao, 1371, CEP 05508-090, S\~ao Paulo, Brazil\\
$^{3}$ Sorbonne Universit\'e, Universit\'e Paris Diderot, CNRS/IN2P3, Laboratoire de Physique Nucl\'eaire et de Hautes Energies, LPNHE, 4 Place Jussieu, F-75252 \\ Paris, France\\
$^{4}$ Aix Marseille Univ, CNRS, CNES, LAM, Marseille, France\\
$^{5}$ Instituto de Astrof\'isica de Andaluc\'ia (CSIC), P.O. Box 3004, 18080 Granada, Spain\\
$^{6}$ Centro de Estudios de F\'isica del Cosmos de Arag\'on (CEFCA), Plaza San Juan, 1, 44001, Teruel, Spain\\
$^{7}$ Departamento de Astrof\'isica, Universidad de La Laguna, E-38206 La Laguna, Tenerife, Spain\\
$^{8}$ Instituto de Astrof\'isica de Canarias, E-38200 La Laguna, Tenerife, Spain\\
$^{9}$ Department of Astronomy, University of Illinois at Urbana-Champaign, Urbana, IL 61801, USA\\
$^{10}$ Donostia International Physics Center, Paseo Manuel de Lardizabal 4, E-20018 Donostia-San Sebastian, Spain\\
$^{11}$ Ikerbasque, Basque Foundation for Science, E-48013 Bilbao, Spain\\
The remaining affiliations can be found after the references.\\
}

\date{Accepted XXX. Received YYY; in original form ZZZ}

\pubyear{2022}

\begin{document}
\label{firstpage}
\pagerange{\pageref{firstpage}--\pageref{lastpage}}
\maketitle

\begin{abstract}
In this series of papers, we employ several machine learning (ML) methods to classify the point-like sources from the miniJPAS catalogue, and identify quasar candidates. Since no representative sample of spectroscopically confirmed sources exists at present to train these ML algorithms, we rely on mock catalogues. In this first paper we develop a pipeline to compute synthetic photometry of quasars, galaxies and stars using spectra of objects targeted as quasars in the Sloan Digital Sky Survey. To match the same depths and signal-to-noise ratio distributions in all bands expected for miniJPAS point sources in the range $17.5\leq r<24$, we augment our sample of available spectra by shifting the original $r$-band magnitude distributions towards the faint end, ensure that the relative incidence rates of the different objects are distributed according to their respective luminosity functions, and perform a thorough modeling of the noise distribution in each filter, by sampling the flux variance either from Gaussian realizations with given widths, or from combinations of Gaussian functions. Finally, we also add in the mocks the patterns of non-detections which are present in all real observations. Although the mock catalogues presented in this work are a first step towards simulated data sets that match the properties of the miniJPAS observations, these mocks can be adapted to serve the purposes of other photometric surveys.
\end{abstract}

\begin{keywords}
quasars: general -- methods: data analysis -- surveys -- catalogues -- techniques: photometric
\end{keywords}



\section{Introduction}

Ongoing and future photometric surveys will gather large data sets across vast volumes, shedding light on our current understanding about cosmology, as well as the formation and evolution of galaxies. Examples of such surveys are the Dark Energy Survey (DES; \citealt{Abbetal}), the Vera C. Rubin Observatory Legacy Survey of Space and Time (LSST; \citealt{Ivezic}), Euclid (\citealt{Amendola:2012ys}), and the Javalambre-Physics of the Accelerated Universe Astrophysical Survey (J-PAS; \citealt{JPAS14}). Automated classification methods are crucial to optimally catalogue and exploit sources observed in large multi-band surveys. In this context, assembling a reliable sample of photometrically selected quasars is a particularly challenging task.

Quasars are extremely luminous active galactic nuclei (AGNs), powered by accretion of matter on to a central supermassive black hole (\citealt{Salpeter64}; \citealt{Zeldovich64}). These astronomical objects are not only the brightest and one of the most highly biased tracers of large-scale structure (\citealt{Porciani04}; \citealt{Croom05}; \citealt{Shen07} \citealt{dAngela08}; \citealt{Ross09}; \citealt{Leistedt13}; \citealt{Leistedt14}; \citealt{Eftekharzadeh15}; \citealt{Laurent17}), but they also share with their host galaxies mutual mechanisms of self-regulatory feedback processes that impact on the galaxy growth, making them a key ingredient in galaxy evolution models (\citealt{Kauffmann2000}; \citealt{DiMatteo05}; \citealt{Schaye15}; \citealt{Sijacki15}; \citealt{Harrison17}). Since they can be seen at large distances, quasars also work as `lighthouses', serving as background light sources to map the intervening neutral hydrogen gas through the Gunn-Peterson effect (\citealt{GunnPeterson}; \citealt{Lynds71}; \citealt{Sargent80}), resulting in the so-called Lyman-$\alpha$ forest.

Their UV-optical spectral energy distributions (SEDs) are characterized by a thermal component from the accretion disk emission in the UV/optical, and a non-thermal continuum from the EUV to the X-rays, a series of broad and/or narrow emission lines\footnote{In the unified model of AGNs (\citealt{Antonucci93}; \citealt{Urry95}), each object receives a different nomenclature depending on the viewing angle to the centre of the source, and the presence of obscuring material. Throughout this paper, we shall use indistinctly ``quasars'' to refer to type-I AGNs -- unless otherwise specified.} (e.g. \citealt{VandenBerk}), blended iron lines (e.g. \citealt{Vestergaard01}; \citealt{VeronCetty04}), as well as a degree of dust reddening at times (e.g. \citealt{Hopkins04}). In photometric images, quasar candidates typically appear as point-like sources and, thus, they can be easily confused with stars and even unresolved galaxies, especially in the low signal-to-noise ratio (S/N) regime. In nearby galaxies hosting low-luminosity AGNs, the signal from the host can also contaminate the (weaker) AGN emission. The contamination of quasars from intervening populations was first identified in the context of the use of broad-band imaging to pre-select spectroscopic targets for the Sloan Digital Sky Survey (SDSS; \citealt{Ross12}; \citealt{Richards09}; \citealt{Leistedt13}; \citealt{Leistedt14}). Since fibers end up allocated only to the brightest, most clearly distinguished quasars, this pre-selection is unable to avoid contamination by other sources in colour-magnitude and colour-colour diagrams, leading to sub-optimal source classification, and a target success rate that changes across the survey footprint.

Fortunately, medium-to-narrow multi-band photometric surveys that continuously cover a large wavelength range, such as ALHAMBRA (\citealt{Moles08}), SHARDS (\citealt{SHARDS}), PAUS (\citealt{Marti14}), Subaru COSMOS 20 project (\citealt{Taniguchi15}), J-PLUS (\citealt{Cenarro19}) and S-PLUS (\citealt{SPLUS}), can help to break degeneracies in the quasar identification by resolving some of the broad emission lines of type-I quasars (as well as most broad absorption line objects), and detecting some of the narrow-lines of type-IIs. In particular, \citet{Abramo12} showed that J-PAS (\citealt{JPAS14}) will observe nearly $\sim$ 240 quasars per square degree for a limiting magnitude of $g<23$, which means that a JPAS-like survey of quasars could be the largest and most complete in the redshift range $0.5 \lesssim z \lesssim 4.0$. This creates a a significant potential for probing cosmological phenomena, such as baryon acoustic oscillations and redshift space distortions. This in turn is potentially impactful for the study of dark energy models, modified gravity models (e.g. \citealt{Abramo20}), as well as primordial non-Gaussianities and relativistic effects (e.g. \citealt{Abramo2017}).

Prior to the arrival of the final scientific instrument (JPCam; \citealt{JPCam1}; \citealt{JPCam2}), the J-PAS telescope (JST/T250) was equipped with a single CCD camera, called JPAS-Pathfinder, which carried out the first observations in nearly 1 deg$^{2}$ on the All-wavelength Extended Groth strip International Survey (AEGIS) field, and tested the performance of the J-PAS optical system. This science verification survey, dubbed miniJPAS (\citealt{miniJPAS}), is a proof of concept for the forthcoming J-PAS survey, allowing us to test the precision with which J-PAS will be able to classify sources, and extract the photometric redshifts of galaxies and quasars.

In addition to the science that J-PAS will be able to conduct using the quasars it identifies, the collaboration is joining efforts with the WEAVE survey (\citealt{WEAVE}), a multi-object spectrograph that will start observing in 2022. Part of the WEAVE strategy is to follow-up high-redshift ($z \geq 2.1$) quasars to conduct a Lyman-$\alpha$ forest and metal line absorption survey (\citealt{Pieri}). The targets for this WEAVE-QSO survey will be provided mainly by J-PAS, which is currently the only instrument capable of identifying quasars in numbers, and down to the depths needed by WEAVE-QSO to do its science.

Because of their high interest to both cosmology and galaxy evolution, the task of building a complete sample of quasar targets is more pressing than ever. Besides, in this new era of massive data acquisition we need effective statistical methods to classify the millions of sources that will be detected by these multi-band photometric surveys, and e.g. identify the quasar candidates. This can be accomplished by using two approaches (or even a combination of both): template-fitting methods (e.g. \citealt{Eldar}) and machine learning algorithms (ML; e.g. \citealt{Golob21}; \citealt{Gines21}; \citealt{Nakazono21}).

Recently, ML approaches have become a powerful tool in astronomy, being preferred when dealing with massive data sets. However, this comes at the expense of requiring complete and representative training sets to ensure that the distribution of main features for each class of astronomical object will be reflected in the resulting trained models, and will be recovered with a good predictive performance on the test sets. For some ML applications in the context of source classification see e.g. \citet{Odewahn92}; \citet{Odewahn2004}; \citet{Fadely12};  \citet{Sevilla-Noarbe18}; \citet{Cabayol19}.

In particular, \citet{Baqui20} employed several different ML methods to perform a star/galaxy separation in the miniJPAS catalogue using both photometric and morphological information, as well as spectroscopically confirmed miniJPAS sources for the training. To go beyond this morphological classification scheme requires a three-class separation (see e.g. \citealt{Ball2006}; \citealt{Brescia15}; \citealt{Clarke2020}; \citealt{Nakazono21}) to have a more thorough assessment of the contaminants in the miniJPAS quasar sample.

However, we lack a representative sample of spectroscopically confirmed sources in the area surveyed by miniJPAS. In particular, only a few hundred spectroscopic quasars were observed in that region. In fact, even if all the sources in miniJPAS had perfect types and redshifts, that is still fall far of the numbers needed to train ML methods. Furthermore, it is not clear whether there will be, in the near future, a sufficiently large and sufficiently deep photometric catalogue of objects with secure (spectroscopic) classification. All this motivated us to develop realistic mock catalogues for use until such a time when both our photometric data and spectroscopic follow-up data reach sufficient size.

Generating synthetic fluxes allows us to asses some basic properties of data sets from upcoming surveys, such as selection effects, the uncertainties in derived galaxy properties, and the relative impact of the different sources of errors. They further allow us to make forecasts exploring survey strategies. Hence, our mock catalogues are crucial for assessing the quality of the miniJPAS classification, and for serving our purposes during the initial phases of the J-PAS survey.

In this paper we describe the methodology adopted for generating simulated fluxes of quasars, galaxies and stars which are based on the properties of the miniJPAS observations. These mock catalogues will be employed for training and validating the performances of several ML algorithms, classifying the miniJPAS point-like sources, and identifying quasar candidates. The ML codes employed in the classification, as well as their individual and combined performances (on test sets of both simulated fluxes and real observations), will appear in subsequent papers (Rodrigues et al. in prep.; Mart\'inez-Solaeche et al. in prep.; P\'erez-R\`afols et al. in prep. 2022a). These classifiers provide scores which can be used as probabilities that any given object is a quasar (at low $z<2.1$ or high $z\geq 2.1$ redshift), a star or a galaxy. In P\'erez-R\`afols et al. in prep. (2022b) we also present a primary catalogue of miniJPAS quasar candidates, which will be more thoroughly investigated with spectroscopic follow-up in the future.

This paper is organized as follows. Section \ref{sec:Data} outlines the miniJPAS and SDSS data employed in this work, and describes the sample selection criteria. In Sect. \ref{sec:Mocks} we present our pipeline for generating simulated photospectra of quasars, galaxies and stars from SDSS spectra, and describe the luminosity functions and noise models. In Sect. \ref{sec:Results} we validate the mock catalogues, and compare the main properties of the simulated fluxes with the miniJPAS observations. In Sect. \ref{sec:Discussion} we suggest some improvements that could further optimize our mock catalogues, and provide additional applications. Finally, we summarize our main findings in Sect. \ref{sec:Conclusions}. All magnitudes here are presented in the AB system.

\section{Data preparation} \label{sec:Data}

In this section we describe the data sets used in this work, which consist in photometric observations from the miniJPAS catalogue (\citealt{miniJPAS}), and spectra of quasar targets from the SDSS Superset catalogue (\citealt{Paris17}).

\subsection{miniJPAS}
\label{subsec:mJPAS} 

The miniJPAS survey (\citealt{miniJPAS}) imaged 0.895 deg$^{2}$ of the Extended Groth Strip (EGS) in four overlapping pointings. The observations\footnote{All miniJPAS images and catalogues are publicly available through the CEFCA web portal: \textcolor{blue}{\url{http://archive.cefca.es/catalogues/minijpas-pdr201912}}.} were conducted with the full J-PAS photometric system, which consists of 54 narrow-band (NB) filters ranging from 3\,780 {\AA} to 9\,100 {\AA} plus two medium-band filters centred on 3\,497 {\AA} (named $uJAVA$) and 9\,316 {\AA} (named $J1007$), and complemented with three SDSS-like broad-bands ($uJPAS$, $gSDSS$ and $rSDSS$). The 54 NB filters present full widths at half maximum (FWHM) of $\sim 145$ {\AA} and are equally spaced every $\sim 100$ {\AA}, whereas the FWHM of the $uJAVA$ and $J1007$ bands are 495 {\AA} and 635 {\AA}, respectively. In addition to these filters, the miniJPAS observations also included the $iSDSS$ broad-band (in a total of 60 filters).

The miniJPAS data were calibrated and reduced by the Data Processing and Archiving Unit at CEFCA (\citealt{UPAD14}). In our analyses, we included only data from the primary catalogue (PDR201912), which contains 64\,293 sources with detection in the $r$-band. The photometry for all sources in this catalogue was obtained with \texttt{SExtractor} (\citealt{SExtractor}) in the dual-mode configuration, and the different types of apertures were defined using the $r$-band as the reference filter. For further details about the observations and data reduction, see \citet{miniJPAS}.

\subsection{SDSS Superset}
\label{subsec:SDSS_Superset} 

The first step to create the mocks was to assemble a large sample of objects with reliable classification, redshifts, and optical spectra, that covered most of the J-PAS wavelength range. 
Fortunately, the area observed by the miniJPAS survey has been observed by a wealth of multi-wavelength facilities, such as AEGIS (\citealt{Davis07}), ALHAMBRA (\citealt{Moles08}), DEEP2/DEEP3 (\citealt{DEEP3a}; \citealt{DEEP3b}; \citealt{DEEP2}), SDSS (\citealt{Dawson13}), and HSC-SSP (\citealt{Aihara18}, \citeyear{Aihara19}). We opted to construct the synthetic photospectra from the publicly available dataset SDSS DR12Q Superset\footnote{The SDSS DR12Q Superset catalogue is available at: \\ \textcolor{blue}{\url{https://data.sdss.org/sas/dr12/boss/qso/DR12Q/Superset\_DR12Q.fits}}} (\citealt{Paris17}) which contains all quasar targets from the final data release of the Baryon Oscillation Spectroscopic Survey (BOSS; \citealt{Dawson13}), as part of the SDSS-III Collaboration (\citealt{Eisenstein11}). The SDSS DR12 spectra were obtained by the BOSS spectrograph, which covers the wavelength region 3\,600-10\,000 {\AA} at a spectral resolution in the range $1\,500<R<2\,500$, using 2$\arcsec$-diameter fibers.

This superset of visually inspected spectra and redshifts provides a census of not only quasars, but also stars and galaxies whose broad-band colours are consistent with those of quasars. Therefore, the combination of visual inspection plus highly-specialized targeting algorithms makes the SDSS Superset catalogue ideal to select not only spectroscopically confirmed quasars, but also the main contaminants in the quasar sample. In particular, the Superset catalogue constitutes a reliable starting point for generating the mocks.

Although some quality metrics come from SDSS DR16 (\citealt{Ahumada20}) with its pipeline refinements, here we limit ourselves to SDSS DR12 (\citealt{Dawson13}), because sample veracity is of critical importance for the classification of miniJPAS sources, and SDSS DR12Q spectra have all been visually inspected. All the analyses described in this paper can be trivially applied to the latest SDSS data release.

\subsection{Sample selection}
\label{subsec:Sample_selection}

\subsubsection{Superset original sample} \label{subsubsec:superset_orig_sample}

The SDSS DR12Q Superset contains 546\,856 quasar targets. For the purposes of this work, we selected spectra that satisfied the following criteria: \texttt{zWARNING}=0 (good-quality spectra); \texttt{SN\_MEDIAN\_ALL}>0 (further quality information from the median signal-to-noise ratio per resolution element); \texttt{Z\_CONF}<3 (large confidence rating for the visually inspected redshift); \texttt{CLASS\_PERSON}=1, 3, 30, and 4 (object classification via visual inspection as star, quasar, broad-absorption line quasar and galaxy, respectively), and apparent magnitudes in the range $17.5 \leq r < 24$. Note, however, that the spectra available for galaxies are limited to the range $18.7\leq r < 24$. This results in a sample of 281\,208 quasars at $z<4.3$, 20\,021 galaxies at $z<0.9$, and 131\,430 stars -- including main sequence stars, white dwarfs (WD), carbon stars (C), and cataclysmic variables (CV).

From Data Release 2 and beyond, the SDSS final calibrated spectra are not corrected for Galactic dust reddening. We correct for Galactic extinction using Milky Way dust maps from \citet{Schlegel98}, who combined results of IRAS (\citealt{IRAS}) and COBE/DIRBE (\citealt{COBE}; \citealt{DIRBE}). The extinction coefficients are obtained from the IRSA Dust Extinction Service using the \texttt{astroquery} module: \footnote{\textcolor{blue}{\url{https://astroquery.readthedocs.io/en/latest/ipac/irsa/irsa_dust/irsa_dust.html}}.}, and we follow the extinction law from \citet{Fitzpatrick07}.

\subsubsection{Superset faint sample} \label{subsubsec:superset_faint_sample}

The SDSS-III/BOSS survey selected quasar candidates by various selection algorithms -- for a detailed description of the BOSS quasar target selection, see \citet{Ross12}. This selection is designed to be sensitive to quasars in the range $2.15 < z < 3.5$, and is limited to $r \lesssim 21.85$. To construct a fair sample of simulated fluxes that resembles the luminosity properties from the miniJPAS observations and reaches $r \sim 24$, we need thus a parent sample of spectra which includes, on average, many more faint sources than the original Superset catalogue does. Following \citet{Abramo12}, we adopt a procedure that consists in building an augmented set of objects by generating new number counts with randomly fainter magnitudes while preserving their spectroscopic identification (i.e. redshift and type). In the case of quasars and galaxies, from each spectra we generate $\sim$20 new fainter objects per magnitude bin, where we consider bins of size 0.5 mag; for stars, this number can vary from 40 to 120 new objects per bin (depending on the spectral type). 

In performing this operation we assume a weak spectral dependence on apparent magnitude (or, equivalently, luminosity). This might not always be true in the case of AGNs due to the Baldwin effect (\citealt{Baldwin77}), the anti-correlation between the continuum luminosity and the rest-frame equivalent widths of UV emission lines (such as Ly$\alpha$ and CIV). This assumption of no luminosity evolution might also be not very realistic for non-active galaxies in general, given that at the same redshift, fainter galaxies tend to be less massive; and thus younger, bluer, and with stronger emission lines. This effect is partially mitigated by the fact that our selection of galaxy spectra was obtained independently of the spectral type (see Sect. \ref{subsubsec:GLF}), i.e. without explicitly considering the relative frequencies of red and blue galaxies separately. Therefore, although the current version of the galaxy mocks might not be the most appropriate one for e.g. galaxy evolution studies, for the moment it corresponds to the best available sample of main contaminants to the miniJPAS quasar population.

\begin{figure}
\includegraphics[width=\linewidth]{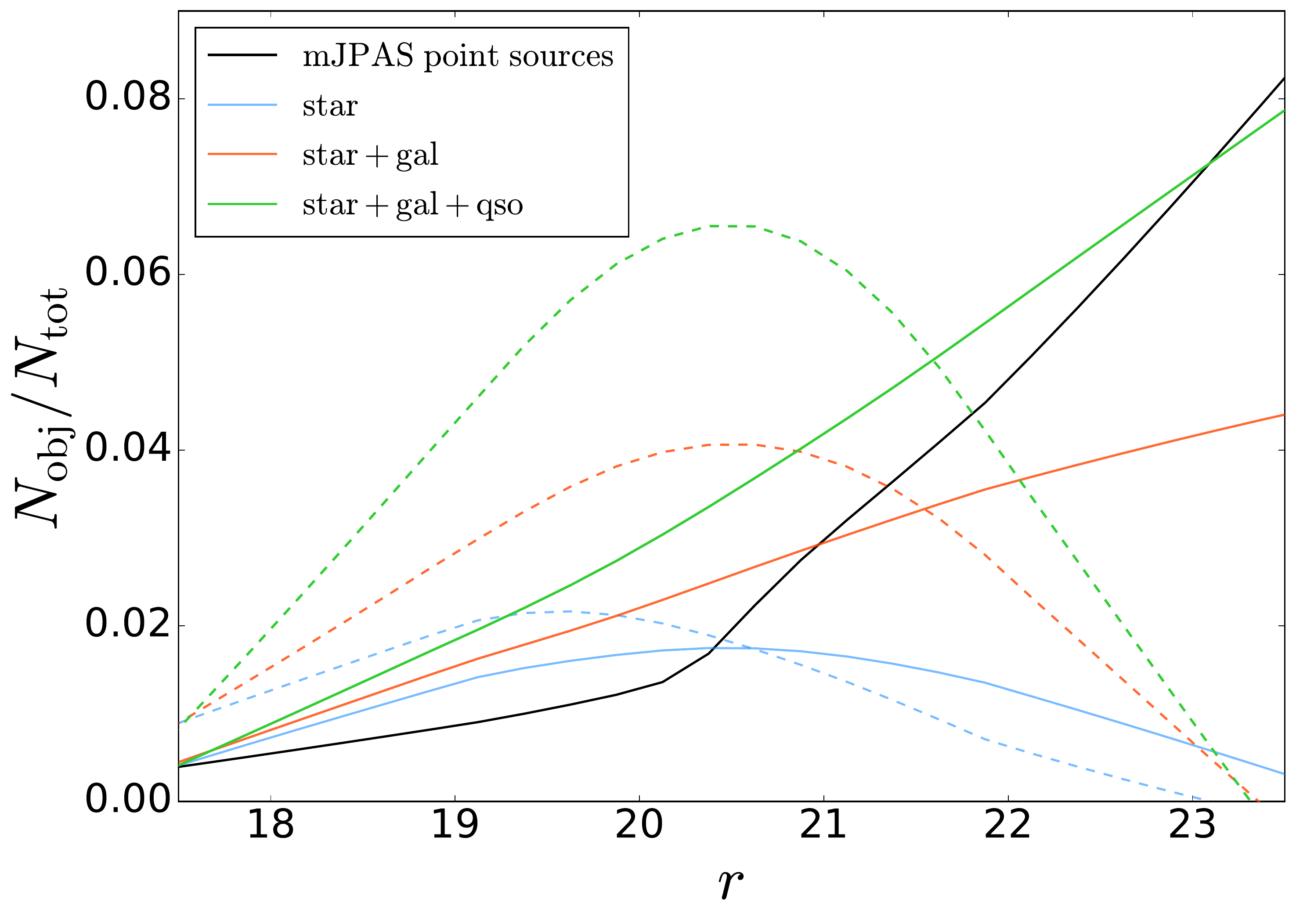}
\caption{Original (dashed lines) and shifted (solid coloured lines) SDSS $r$-band magnitude distributions of a balanced sample of 30\,000 spectra of stars, galaxies and quasars. As a comparison, the black solid line shows the distribution of the miniJPAS point-like sources. The y-axis corresponds to the number density of objects per magnitude bin.}
\label{fig:mag_comparison}
\end{figure}

In Fig. \ref{fig:mag_comparison} we illustrate how the original SDSS $r$-band magnitude distributions are shifted to fainter magnitudes (by fixing their redshifts and stellar types) in a balanced sample of 30\,000 spectra of stars, galaxies and quasars. As we can see, the assigned fainter magnitudes yield a more representative distribution of the miniJPAS point-like sources (see Sect. \ref{subsubsec:miniJPAS_point_sample}).

Finally, we map the augmented set of spectra into the number counts specified by the corresponding putative luminosity functions (described in Sect. \ref{subsec:LF}). The objects selected in this way constitute our final sample of parent spectra. This mapping is illustrated in Fig. \ref{fig:magr_z_gal} for the galaxies.

\subsubsection{miniJPAS sample} \label{subsubsec:photo_sample}

The miniJPAS catalogue contains some quality cut flags warning whether each source has an issue in one or more filters that may impair or invalidate the photometry. The $\texttt{FLAGS}$ column comprises the $\texttt{SExtractor}$ flags and indicates close neighbors, blending, saturation, truncation, and so on. The $\texttt{MASK\_FLAGS}$, in turn, informs if the object is outside the window frame, whether it is a bright star or is located near one, and if it has a nearby artifact. To ensure good photometry we only selected miniJPAS sources with $\texttt{FLAGS}$=0 and $\texttt{MASK\_FLAGS}$=0 in all bands. This so-called non-flagged subsample contains 46\,440 sources with measured photometry in 60 or less bands.

As part of the calibration process (\citealt{LopezSanjuan2019}), the miniJPAS photometry is already corrected for atmospheric extinction. Hence, we only need to correct the miniJPAS data for Galactic extinction, for which we employ a similar procedure as the one outlined for the SDSS spectra in Sect. \ref{subsubsec:superset_orig_sample}.

The miniJPAS catalogue provides different types of photometric measurements in all the bands. Since our ultimate goal is to classify point-like sources, in our analyses we employ $\texttt{APER\_3}$ magnitudes (i.e. apparent magnitudes computed within a 3$\arcsec$-diameter aperture). This choice is made to guarantee high-accuracy photometric redshifts for quasars, and will be discussed in more detail in an upcoming paper (Queiroz et al. in prep.). Given that a 3$\arcsec$-aperture misses part of the total light emitted by the source, an aperture correction $\Delta m_{\mu(a)}^{3\arcsec}$ is applied to these magnitudes. The correction term per passband per tile is derived by using spectroscopically confirmed non-saturated stars from the SDSS Superset catalogue that were observed by miniJPAS, and it is computed from the $\texttt{APER\_6}$ corrections ($\Delta m_{\mu(a)}^{6\arcsec}$) available in the \texttt{miniJPAS.TileImage} table as follows:

\begin{equation}
    \displaystyle{\Delta m_{\mu(a)}^{3\arcsec}=\mathrm{median} \ \left[m^{6\arcsec}_{\mu(i,a)}+\Delta m_{\mu(a)}^{6\arcsec}-m^{3\arcsec}_{\mu(i,a)}\right]} \; ,
\label{eq:delta_mag_aper}
\end{equation}
where $m_{\mu(i,a)}$ is the observed aperture magnitude of the $i$-th star measured in band $\mu$ in tile $a$.

\begin{figure*}
\includegraphics[width=\textwidth]{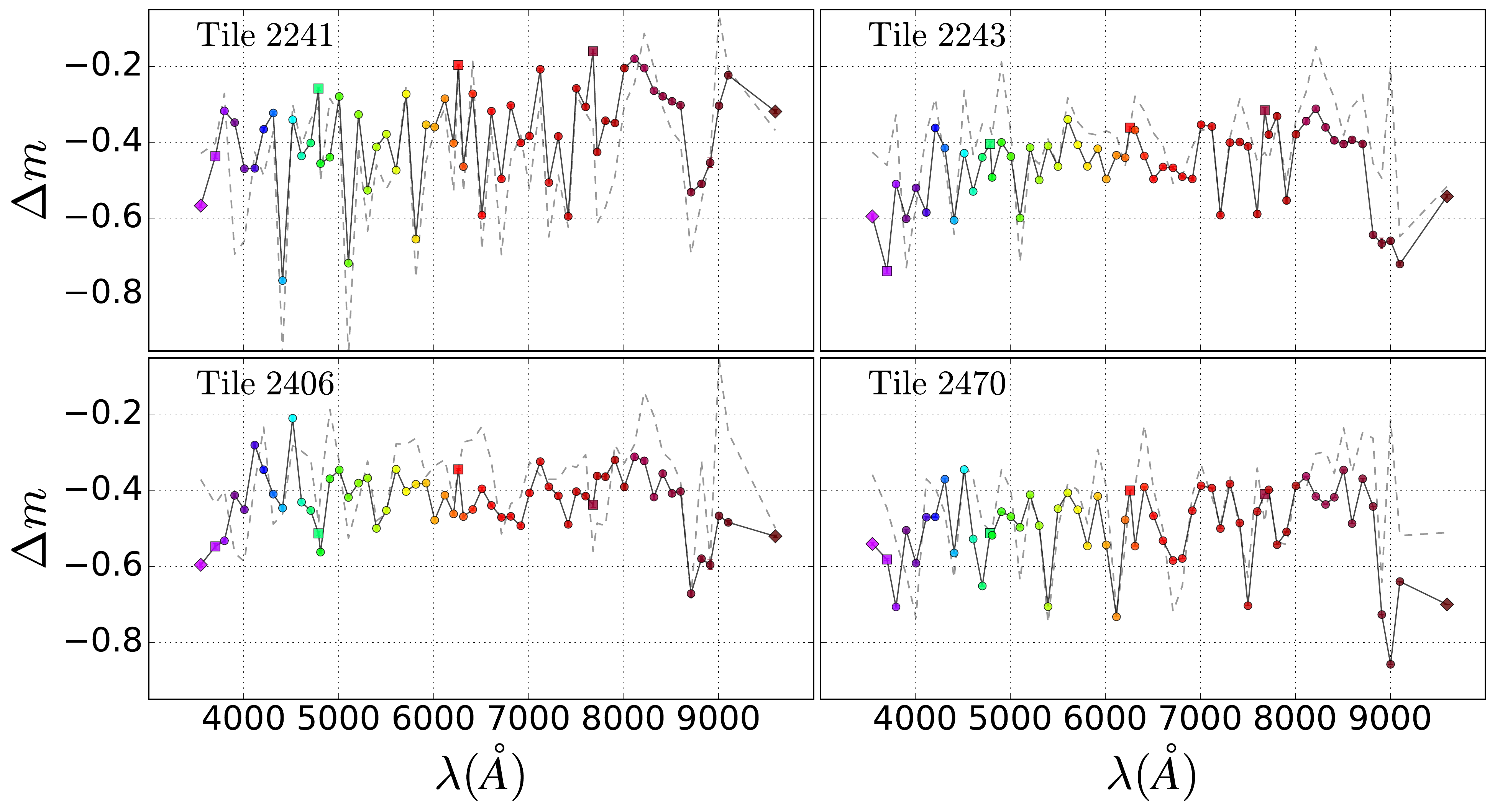}
\caption{Final offsets (coloured symbols) as a function of the tile applied to 3$\arcsec$-aperture magnitudes of the non-flagged miniJPAS sample. Narrow, intermediate and broad-bands are represented by circles, diamonds and squares, respectively. The gray dashed lines correspond to the aperture corrections alone, i.e. $\Delta m_{\mu(a)}^{3\arcsec}$.}
\label{fig:aper_corr}
\end{figure*}

In addition to these aperture corrections, we use the same non-saturated stars to compute the systematic offsets $\delta m_{\mu(a)}$ in the photometry: 

\begin{equation}
    \displaystyle{\delta m_{\mu(a)}=\mathrm{median} \ \left[m_{\mu(i,a)}^{3\arcsec}-m_{\mu(i,a)}^{synt}\right]} \; ,
\label{eq:delta_mag_offset}
\end{equation}
where $m_{\mu(a)}^{synt}$ is the synthetic magnitude in band $\mu$ for the $i$-th star. The final magnitude corrections per passband per tile are then given by 

\begin{equation}
    \displaystyle{\Delta m_{\mu(a)}=\Delta m_{\mu(a)}^{3\arcsec}+\delta m_{\mu(a)}} \; ,
\label{eq:delta_mag_star}
\end{equation}
and are shown in Fig. \ref{fig:aper_corr}. As we can see, these offsets are non-negligible, and can be as large as 0.8 mag in absolute value for some filters.

The miniJPAS observations were performed in groups of seven filters, and carried out with different sky conditions. Moreover, the reddest filters were observed last, when the AEGIS field reached the lowest elevations (i.e. highest air mass measurements). This results in a non-flagged sample containing only 2\,423 sources with observations in all 60 bands. To build realistic mock catalogues, we need to model and include these non-detections (ND) into the simulated fluxes.

In the miniJPAS catalog, non-detections are assigned with magnitude values of $99.0$, and can have two origins: $(i)$ either they correspond to magnitudes fainter than the limiting sensibility of the detector for that specific band (i.e. they have low signal-to-noise ratios), or $(ii)$ they are related to negative fluxes. After confirming that there are no non-observations in the non-flagged sample, as indicated by a positive value of the normalized weight map flag in all bands, we opted to treat these instances of NDs separately by adopting the following conventions for the miniJPAS photometry: 
\begin{enumerate}
    \item if $\mathrm{S/N}<1.25$: $\displaystyle{[m_{\mu(i)}, \, \sigma_{m, \, \mu(i)}]=(99.0, \ m_{\mathrm{\mu, \, lim}}^{3\arcsec, \, 5\sigma})}$, where $\mathrm{S/N}$ is the signal-to-noise ratio in band $\mu$ for the $i$-th source, and $\displaystyle{m_{\mathrm{\mu, \, lim}}^{3\arcsec, \, 5\sigma}}$ is the targeted minimum depth defined in \citet{JPAS14};
    \item if $F_{\lambda, \, \mu(i)}<0$: $[m_{\mu(i)}, \, \sigma_{m, \, \mu(i)}]=(-99.0, \, 99.0)$, where $F_{\lambda, \, \mu(i)}$ is the measured flux in units $\mathrm{erg} \, \mathrm{s}^{-1} \, \mathrm{cm}^{-2} \,${\AA}$^{-1}$.
\end{enumerate}

\subsubsection{miniJPAS spectroscopic sample} \label{subsubsec:mjpas_spec_sample}

After re-processing the data, we build a miniJPAS spectroscopic sample by cross-matching the non-flagged sample with the Superset sample within a radius of 1$\arcsec$ using the \texttt{TOPCAT} software (\citealt{Topcat}). This cross-match resulted in a set of 117 quasars, 40 galaxies and 115 stars.

We have also cross-matched the miniJPAS non-flagged sample with DEEP3 (\citealt{DEEP3a}; \citealt{DEEP3b}), a dedicated spectroscopic campaign focused on the Extended Groth Strip, within a radius of 1.5$\arcsec$. The DEEP3 survey does not cover the whole optical wavelength range (spanning 4\,550–9\,900 {\AA}), and was designed to map galaxies down to a limiting magnitude of $R\sim24.4$. This implies more noisy spectra which cannot be reliably visual inspected (when compared to the SDSS spectra, for instance). In particular, many of the sources classified as AGNs by DEEP3 at low redshifts seem to be actually galaxies with a very small AGN component, low-luminosity type-Is, and even type-IIs. So in the case of quasars we only select DEEP3 sources that are identified as AGNs at redshifts $z \geq 1.5$. In addition, we also apply the following selection criteria: \texttt{ZQUALITY} $\geq 3$ for quasars and galaxies, and \texttt{ZQUALITY} $= -1$ for stars (an indicator of redshift quality); \texttt{RCHI2} $\geq 0.6$ (reduced $\chi^{2}$ square for the redshift fit); \texttt{PGAL} $\geq 0.6$ for galaxies and \texttt{PGAL} $< 0.5$ for stars (a value between 0 and 1 indicates the probability of a source being a galaxy, for unresolved sources; while a value equal to 3 indicates a resolved galaxy); $r$-band magnitude in the range $17.5\leq r <24.0$. The final miniJPAS-DEEP3 sample contains 15 quasars at $1.5<z<3.7$, 8\,779 galaxies at $z\leq1.7$ (6\,514 at $z<0.9$) and 37 stars. 

\subsubsection{miniJPAS point-like sample} \label{subsubsec:miniJPAS_point_sample}

The miniJPAS database provides different complementary methods to estimate the stellarity index of each source (see \citealt{miniJPAS} for more details). In these classifications an index close to one (zero) indicates that the source is likely to be a star (galaxy). In this work, we employ the Extremely Randomized Trees (ERT) machine learning classifier (\citealt{Baqui20}), which uses both morphological and photometric information. To build our set of miniJPAS point-like sources, we select objects from the non-flagged sample that were classified as likely to be stars with a probability $\mathcal{P}_{\mathrm{ERT}}\geq 0.1$. This quality cut can properly separate extended and point sources up to $r\sim 22$. In order to maximize the selection of point-like sources, whenever $\mathcal{P}_{\mathrm{ERT}}=-99.0$ we also consider objects which were classified as likely stars by the stellar-galaxy locus classifier (SGLC; \citealt{LopezSanjuan19b}) with a probability of $\mathcal{P}_{\mathrm{SGLC}}\geq 0.1$. The final miniJPAS point-like subsample contains 11\,419 objects.

The miniJPAS point-like sample is used as a proxy to select realistic S/N distributions in each band, as well as draw the pattern of non-detections to be applied to the mocks. Finally, we also construct three subsamples containing about 10k point sources each, that are randomly selected according to the targeted number counts provided by the luminosity functions of quasars, galaxies and stars (see Sect. \ref{subsec:LF} for more details). These subsamples are designed to provide fairer comparisons with the corresponding test sets.

\section{Mock catalogues} \label{sec:Mocks}

Machine learning (ML) techniques comprehend an ensemble of adaptive learning methods to recognize and predict some sort of pattern within a specific class of objects. On one hand, such approach has the advantage of not making any prior assumptions regarding the types of objects or their evolution; on the other, it requires representative training sets to estimate the learning model and achieve high classification performances.

To classify the miniJPAS point sources, and identify the quasar candidates, we need a large and sufficiently complete sample of spectroscopically confirmed objects with miniJPAS observations. However, in the AEGIS field we fall far short of a sufficiently large sample, and the available one is neither fair nor complete to the depth we require. Consequently, the AEGIS field cannot be used alone to properly train the ML algorithms. 

To bypass this fundamental limitation (which is shared with many forthcoming photometric surveys) we generated mock catalogues of simulated photospectra of quasar targets, and developed a dedicated pipeline to include realistic features from the observations, such as the S/N in all bands, the magnitude-redshift-type distributions drawn from putative luminosity functions, as well as the pattern of non-detections. Ensuring the accuracy of these simulated photospectra is key for the generalization of the models trained on mocks to real data. For this reason, this constitutes a constant work-in-progress, in the sense that, as we acquire more information from observations over larger areas, with larger samples of spectroscopically confirmed sources, we will be able to further refine the mocks.

\begin{figure*}
\includegraphics[width=\textwidth]{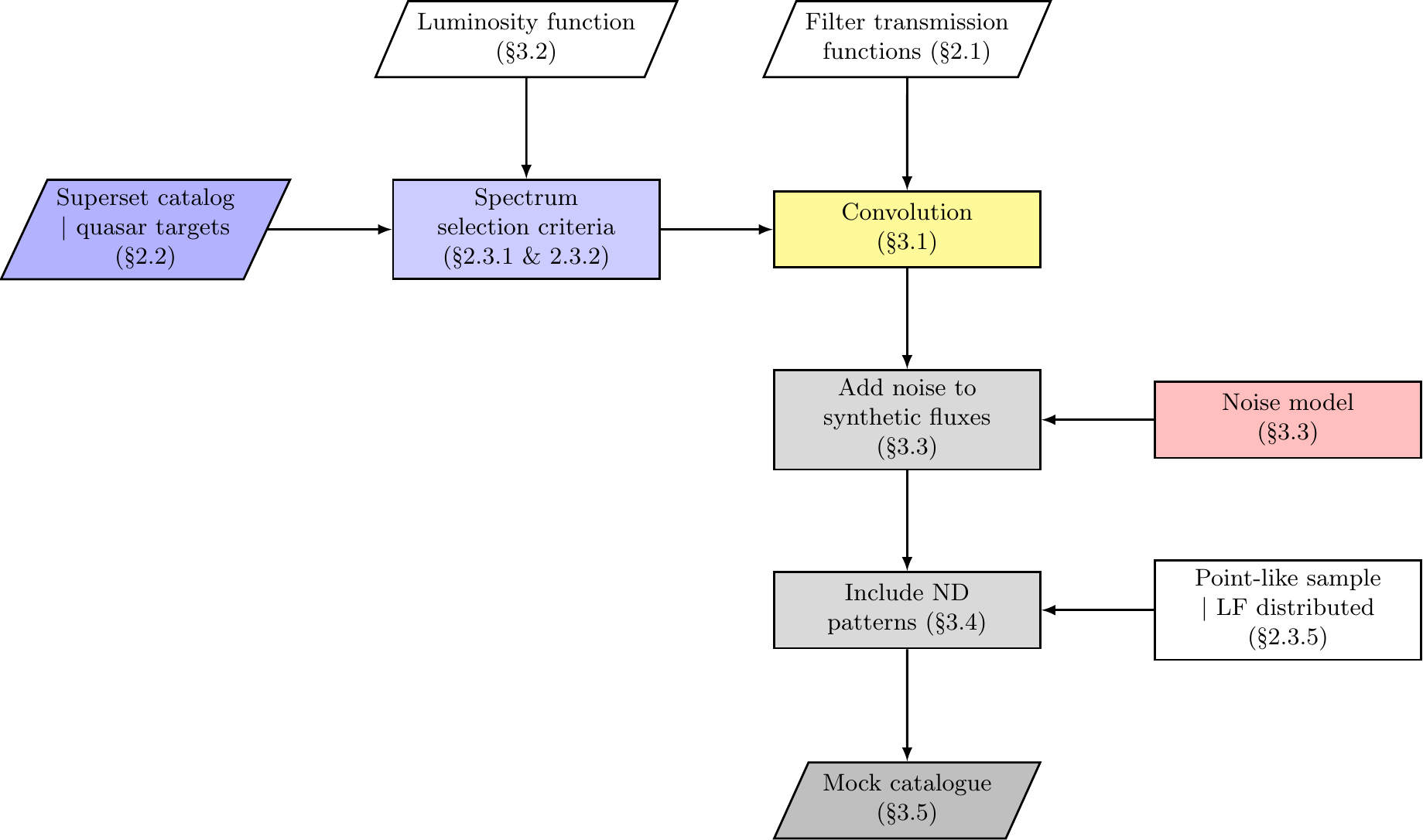}
\caption{Pipeline to simulate photospectra with the same signal-to-noise ratio distributions expected for observations within a given photometric system.}
\label{fig:pipeline}
\end{figure*}

In Fig. \ref{fig:pipeline} we present the methodology developed to build the mock catalogues. Each block contains the reference to the location in the text. The Superset and miniJPAS point-like samples were already discussed in Sections \ref{subsubsec:superset_faint_sample} and \ref{subsubsec:miniJPAS_point_sample}, respectively. Throughout the following subsections we describe in more detail each of the subsequent steps of the pipeline. Although explicitly applied for the J-PAS photometric system here, this procedure can be easily adapted for any other multi-band optical survey.

\subsection{Synthetic photometry} \label{subsec:synt_flux}

The synthetic photometry is computed using the prescription described in \citet{DiazGarcia15}, based on the HST \texttt{synphot}\footnote{\textcolor{blue}{\url{http://stsdas.stsci.edu/Files/SynphotManual.pdf}}} package, and in \citet{Bessel05}; \citet{Pickles2010}. 

Briefly, for modern photon-counting devices the synthetic fluxes (in unit wavelength) $F_{\lambda,\mu(i)}^{synt}$ can be obtained by
\begin{equation}
    \displaystyle{F_{\lambda, \, \mu(i)}^{synt}=\frac{\int T_{\mu}(\lambda) \, S_{i}(\lambda) \, \lambda \, d\lambda}{\int T_{\mu}(\lambda) \, \lambda \, d\lambda}} \; ,
\label{eq:synt_flam}
\end{equation}
where $S_{i}(\lambda)$ is the dereddened SED of the $i$-th source, and $T_{\mu}(\lambda)$ is the total efficiency of the transmission curve of filter $\mu$. The SED is scaled to match the $r$-band SDSS photometry, which can be either the original value or a new randomly assigned fainter one. 

ST magnitudes (i.e. per unit wavelength) are defined as

\begin{equation}
    \displaystyle{m_{\mathrm{ST}, \, \mu(i)}^{synt}=-2.5\log_{10}F_{\lambda, \, \mu(i)}^{synt}-21.1} \; ,
\label{eq:stmag}
\end{equation}
and AB magnitudes can then be obtained from

\begin{equation}
    \displaystyle{m_{\mathrm{AB}, \, \mu(i)}^{synt}=m_{\mathrm{ST}, \, \mu(i)}^{synt}-5\log_{10}\lambda_{\mathrm{pivot}, \, \mu}+18.692} \; ,
\label{eq:mag_ab}
\end{equation}
where the pivot wavelength is defined as

\begin{equation}
    \displaystyle{\lambda_{\mathrm{pivot}, \, \mu}=\sqrt{\frac{\int{T_{\mu}(\lambda) \, \lambda \, d\lambda}}{\int{T_{\mu}(\lambda) \, d\lambda / \lambda}} }} \; .
\label{eq:lambda_pivot}
\end{equation}

Hereafter, we shall use indistinctly $m_{\mu(i)}^{synt}$ to refer to (synthetic) AB magnitudes. Finally, we assume that the synthetic photometry is centred at the corresponding effective wavelength, given by

\begin{equation}
    \displaystyle{\lambda_{\mathrm{eff}, \, \mu}=\frac{\int{T_{\mu}(\lambda) \, \lambda \, d\lambda}}{\int{T_{\mu}(\lambda) \, d\lambda}} } \; .
\label{eq:lambda_eff}
\end{equation}

Equivalently, the variance of the synthetic flux in each passband can be obtained by error propagation of Eq. \ref{eq:synt_flam}. Note, however, that the median SDSS S/N per pixel scales as $1\sqrt{N_{bin}}$ of the median synthetic S/N, where $N_{bin}\sim 145/0.87$ corresponds to the number of spectral bins that compose the flux in a given narrow-band, considering that the BOSS spectra are measured in spectral bins of $\sim$ 0.87 {\AA}. This implies that the medium SDSS S/N is negligible when compared to the synthetic S/N, and therefore is neglected in the final photometry.

To reach similar signal-to-noise ratio distributions as those of miniJPAS point sources, we still need to add noise to the synthetic photometry, such that the final level of noise in the mocks is similar to the miniJPAS observations. This procedure is described in Sect. \ref{subsec:Noise_model}.

Another caveat concerns the wavelength range spanned by the SDSS DR12 spectra; since they do not fully cover $uJAVA$ and $J1007$, this could preclude us from including those bands in the mocks. Thus, prior to the convolution shown in Eq. \ref{eq:synt_flam} we perform a template fitting of the spectra, in order to extend their coverage. Our method consists in fitting the blue and the red parts of each spectrum separately to ensure a smoother transition when concatenating the fluxes of the two best-fitting templates with the spectrum.

The best-fitting template is chosen by minimizing the following function:

\begin{equation}
    \displaystyle{\chi_{k(i)}^{2}=\sum_{\mu}\left[ \frac{F_{\lambda, \, \mu(i)}^{synt}-\mathcal{T}_{\lambda, \, \mu(k)}}{\sigma_{F, \, \mu(i)}^{sdss}} \right]^{2}}
\label{eq:chi2_template}
\end{equation}
where $\mathcal{T}_{\lambda, \, \mu(k)}$ is the synthetic flux of the $k$-th template scaled by the first blue (or red) filter with a valid observation, and $\sigma_{F, \, \mu(i)}^{sdss}$ is the uncertainty of the synthetic flux (i.e. SDSS noise)\footnote{Note that this is the only step where we have explicitly made use of the flux uncertainties coming from the spectral bins.}. For galaxies, the flux densities of the templates are computed at the spectroscopic redshift.

In the case of quasars, we adopt a single Vanden Berk composite spectrum (\citealt{VandenBerk}) to which we add an adjustable amount of extinction, following the prescription of \citet{HC16}. In the case of galaxies and stars, the best-fitting templates are chosen from a library of SED models (in a total of 37 and 154, respectively) available with the code LePhare (\citealt{Arnouts}; \citealt{Ilbert}).

\subsection{Luminosity function} \label{subsec:LF}

An essential element in our mock catalogues is to build balanced training, validation and test sets containing representative relative frequencies of quasars, galaxies and stars, drawn from putative luminosity functions. For quasars and galaxies, we assume that these number densities depend both on redshift and magnitude, without making any further assumptions on the frequencies of their sub-types. In the case of stars, besides the dependence on magnitude, their number densities also have an angular dependence, but are of course independent of redshift. In addition, since some stellar types, such as A, F, M and white dwarfs, are more often confused with quasars, either due to their similar colours (\citealt{Richards02}) or because their continuum emission can be confused with the Lyman-break of high-redshift quasars, we also considered a dependency on the stellar spectral types. 

Since the effective area surveyed by miniJPAS is small (0.895 deg$^{2}$), to have representative samples to train and assess the performance of the ML classifiers, we simulate multiple realizations of the relative frequencies expected in $\sim$1 deg$^{2}$ until we obtained the desired size for the test, validation and training sets. This procedure is complementary to the one employed to generate more faint sources (described in Sect. \ref{subsubsec:superset_faint_sample}). For instance, the training sets (which contain 100k objects each) correspond to realizations inside final areas of approximately 196, 16, and 46 deg$^{2}$ for quasars, galaxies, and stars, respectively. 

Throughout this paper, we shall use indistinctly luminosity function (LF) to refer to the number counts in luminosity and type (or redshift, if applicable) for each class of object. The corresponding LFs are described in the following subsections.

\subsubsection{Quasar luminosity function} \label{subsubsec:QLF}

For quasars, we adopt the pure luminosity evolution (QLF, hereafter) function from \citet{QLF}, which assumes that the luminosity of all quasars scales up according to some function of redshift, and it allows the bright-end and faint-end slopes to be different on either side of a pivot redshift set as $z_{\mathrm{pivot}} = 2.2$. Considering a perfect selection of objects, we find that over an area of $1/5$ of the full sky  (similar to what is planned for the entire J-PAS footprint), a flux-limited ($r < 23.5$) survey could yield more than three million quasars up to $z=6$, which is in accordance with the estimates from \citet{Abramo12}.

\begin{figure}
\includegraphics[width=\linewidth]{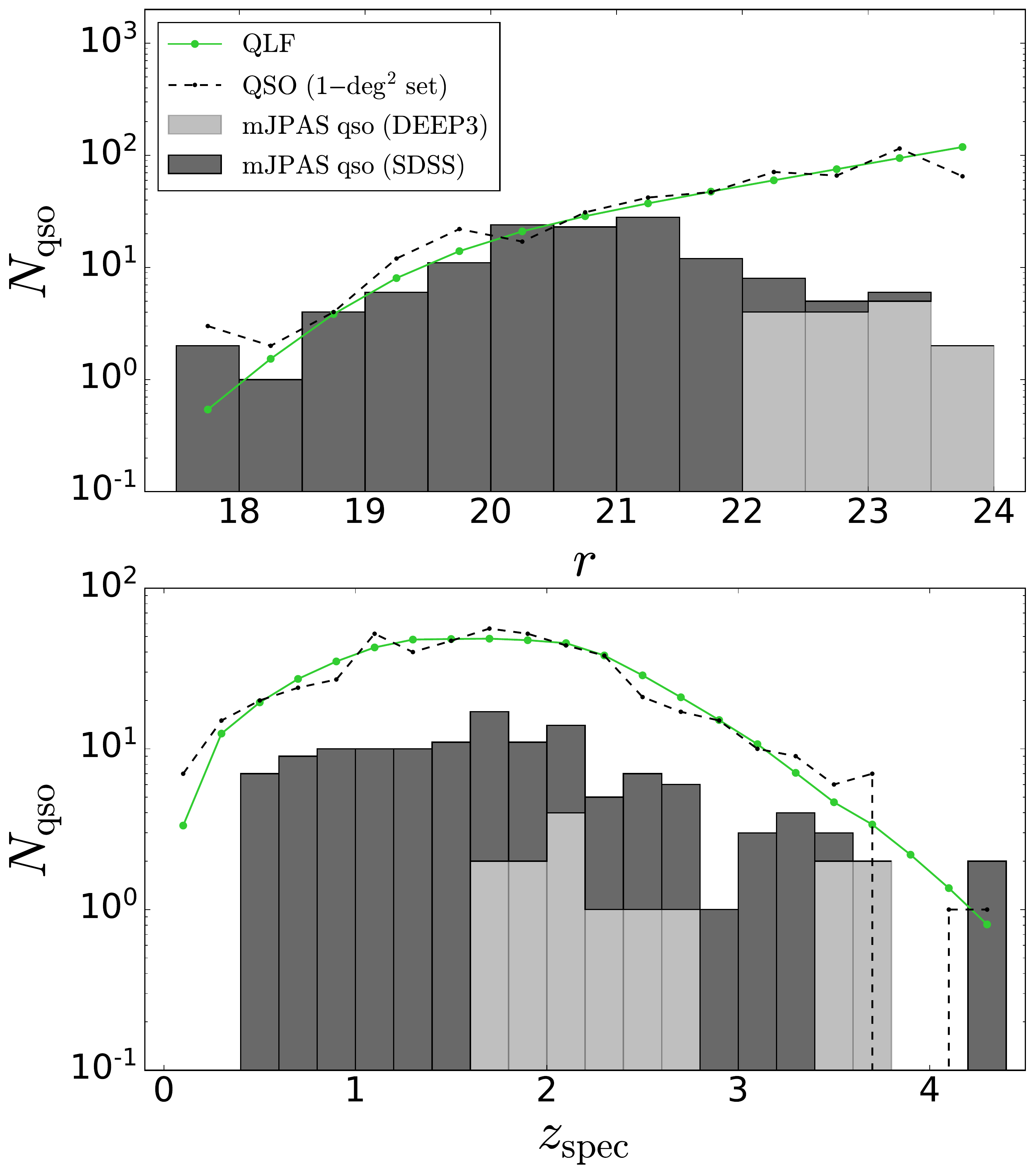}
\caption{Number of quasars per deg$^{2}$ from the luminosity function (green solid line) and miniJPAS (gray bars) as a function of the $r$-band magnitude (top) and spectroscopic redshift (bottom). The miniJPAS quasars from DEEP3 (SDSS) are shown in light (dark) gray. As a comparison, we also show the distribution of quasars in the 1-deg$^{2}$ set (black dashed line) using noise model 11 as reference.}
\label{fig:lf_qso}
\end{figure}

To generate the mock catalogues, we selected quasar spectra in the redshift range $0.033\leq z\leq4.3$ and with magnitudes between $17.5\leq r <24$. In Fig. \ref{fig:lf_qso} we show the magnitude-redshift distribution per deg$^{2}$ predicted by the QLF, and compare it with the distribution of the miniJPAS quasars, separating the contributions from the cross-matches with DEEP3 and SDSS Superset. The QLF predicts 510 quasars per deg$^{2}$, being 133 at high redshifts ($z \geq 2.1$). Although DEEP3 quasars are complementary to the SDSS sample in the faint end, we can still clearly see that the sample of spectroscopically confirmed miniJPAS quasars becomes highly incomplete at $r \gtrsim 21.5$.

In Fig. \ref{fig:lf_qso} we also provide the distributions of quasars in the 1-deg$^2$ set using noise model 11 as reference (see Sect. \ref{subsec:Noise_model} for more details). The absence of quasars at $3.6<z<4.0$ is attributed to cosmic variance.

\subsubsection{Galaxy luminosity function} \label{subsubsec:GLF}

For galaxies, instead of using a phenomenological luminosity function, we mapped the magnitude-redshift distributions of the miniJPAS spectroscopic galaxies. To ensure a fair distribution of galaxies in the bright and faint ends, we combined the contributions from both Superset and DEEP3 galaxies observed with miniJPAS.

\begin{figure}
\includegraphics[width=\linewidth]{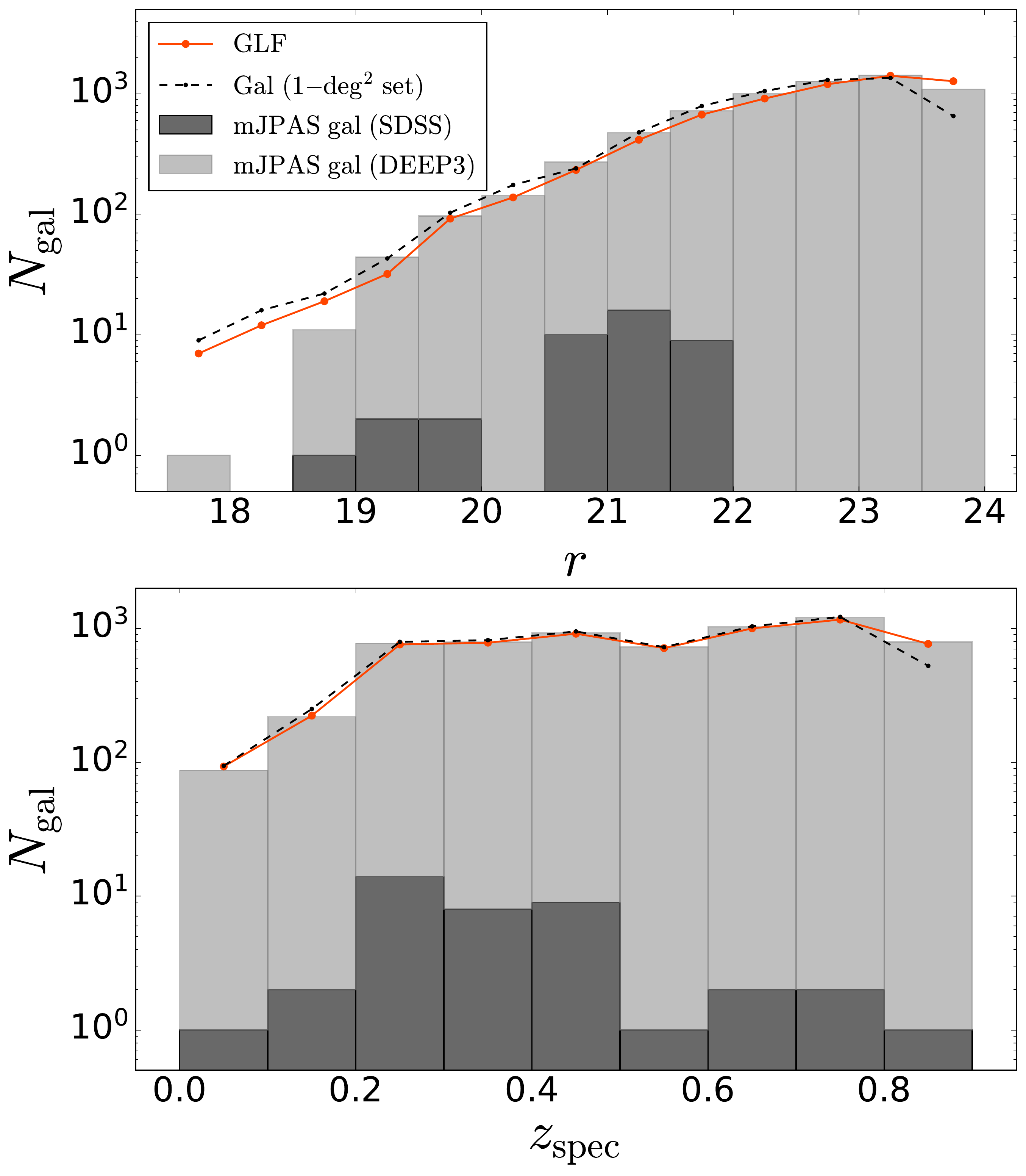}
\caption{Number of galaxies per deg$^{2}$ from the luminosity function (orange solid line) and miniJPAS (gray bars) as a function of the $r$-band magnitude (top) and spectroscopic redshift (bottom). The miniJPAS galaxies from DEEP3 (SDSS) are shown in light (dark) gray. As a comparison, we also show the distribution of galaxies in the 1-deg$^{2}$ set (black dashed line) using noise model 11 as reference.}
\label{fig:lf_gal}
\end{figure}

To generate the mock catalogues, we consider spectra of galaxies at $z\leq0.9$, and with magnitudes between $18.7\leq r <24$. Note that these magnitude-redshift ranges are limited by the availability of spectra in the Superset catalog, and a more detailed separation in the number counts of blue and red galaxies is beyond the scope of this paper. Such galaxy luminosity function (GLF) predicts 6\,410 galaxies per deg$^{2}$. In Fig. \ref{fig:lf_gal} we show the magnitude-redshift distributions predicted by the GLF, and compare them with the distributions of the miniJPAS galaxies. As we can see, the GLF is dominated by galaxies from DEEP3. We also compare the distributions of galaxies in the 1-deg$^{2}$ set using noise model 11 as reference.

\subsubsection{Stellar luminosity function} \label{subsubsec:SLF}

Although the number and types of galaxies are more or less uniformly distributed across the sky, this is not true for stars: their number densities and spectral types are highly dependent on the line of sight that we are probing throughout the Milky Way.

To take this effect into account, we make use of the Besan\c{c}on model of stellar population synthesis of the Galaxy (\citealt{Besancon}) to compute the stellar counts per spectral type per magnitude bin in the same angular position of the miniJPAS area. We consider main sequence stars, white dwarfs, carbon stars and cataclysmic variables in the magnitude range $17.5 \leq r < 24$. These number counts are complemented by the number densities of miniJPAS spectroscopic stars, and we also extrapolate the relative frequencies of the following types: O, B, A, WD, C, and CV. 

\begin{figure}
\includegraphics[width=\linewidth]{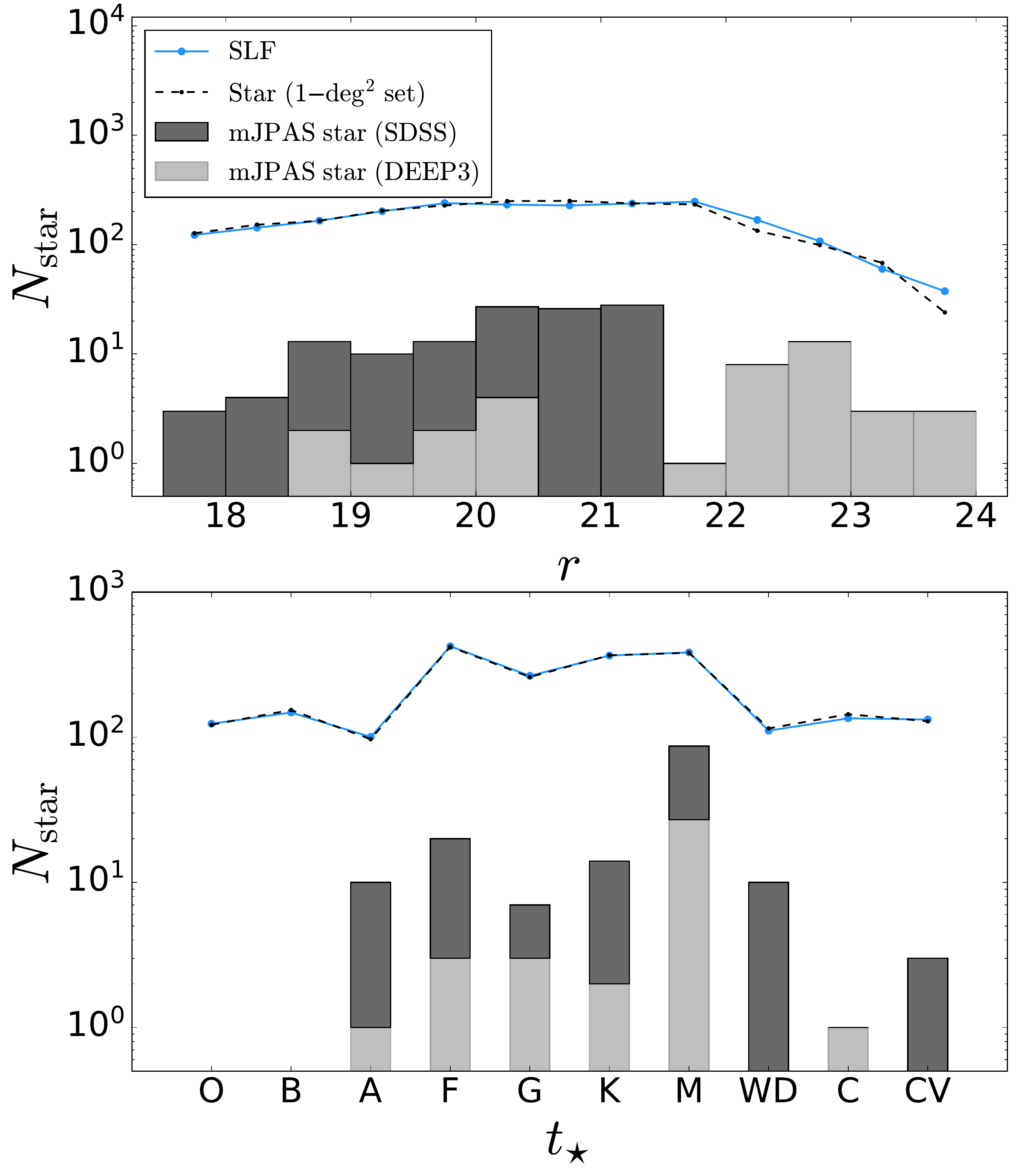}
\caption{Number of stars per deg$^{2}$ from the luminosity function (blue solid line) and miniJPAS (gray bars) as a function of the $r$-band magnitude (top) and spectral type (bottom). The miniJPAS stars from DEEP3 (SDSS) are shown in light (dark) gray. As a comparison, we also show the distribution of stars in the 1-deg$^{2}$ set (black dashed line) using model 11 as reference.}
\label{fig:lf_star}
\end{figure}

Such star luminosity function (SLF) predicts 2\,190 stars per deg$^{2}$ in the AEGIS field. In Fig. \ref{fig:lf_star} we show the magnitude-type distribution of stars predicted by the SLF, and compare it with the distribution in miniJPAS and in the 1-deg$^{2}$ set. As we can see, the spectroscopic sample of miniJPAS stars is highly incomplete in comparison with the distributions expected from SLF. As we show in Mart\'inez-Solaeche et al. (in prep.), this incompleteness is actually an effect of the overestimation of the number densities of some stellar types (especially the more massive ones) in the SLF.

\subsection{Noise model} \label{subsec:Noise_model}

To match similar S/N distributions in all bands as the ones obtained for miniJPAS point sources, we perform a thorough modeling of the noise profiles of the observations, and include realistic errors into the synthetic fluxes derived in Eq. \ref{eq:synt_flam}, as described in the following.

First we sort in a consistent way the magnitudes and fluxes of the miniJPAS point sources in ascending order for each passband. Then, we search in this sorted list for a magnitude value similar to $m_{\mu(i)}^{synt}$, and associate the corresponding magnitude and flux uncertainties (denoted as nominal errors) to the synthetic photometry. The nominal errors in units of flux will be referred as $\sigma_{\mu}$. Given that the nominal errors are associated to the synthetic photometry in a random way, this procedure ensures that the noise patterns from the different tiles are well represented in the mocks. Finally, the flux fluctuation is taken over a realization of a Gaussian function (or a combination of two or more Gaussians) with width proportional to the nominal error. 

We test 10 different noise models, as defined in Table \ref{tab:noise_models}. Such a variety of models help us to ensure a proper modeling of the noise profiles of all bands, even for the faintest miniJPAS sources. Models 1 to 5 correspond to single Gaussian functions with increasing widths; models 6 to 9 correspond to combinations of two or more Gaussian functions with different widths; model 10 samples the red bands ($\lambda_{\mathrm{eff}}\geq7\,416$ {\AA}) with a broader Gaussian than the blue bands; and model 11 corresponds to the best-fitting noise model for each band. 

\renewcommand{\arraystretch}{1.5}
\begin{table}
  \begin{center}
    \caption{Summary of the noise models tested to properly model the miniJPAS observations. Red bands are defined such that $\lambda_{\mathrm{eff}}\geq7\,416$ {\AA}.}
    \label{tab:noise_models}
    \begin{tabular}{c c} 
      \hline
      \hline
      \multicolumn{1}{c}{\textbf{Model}} & \multicolumn{1}{c}{\textbf{Description}}\\
      \hline
      1 & G(0,$ \, 1\sigma_{\mu}$)\\
      2 & G(0,$ \, 1.5\sigma_{\mu}$)\\
      3 & G(0,$ \, 2\sigma_{\mu}$)\\
      4 & G(0,$ \, 2.5\sigma_{\mu}$)\\
      5 & G(0,$ \, 3\sigma_{\mu}$)\\
      6 & $\frac{2}{3}$G(0,$ \, 1\sigma_{\mu}$)+$\frac{1}{3}$G(0,$ \, 2\sigma_{\mu}$)\\
      7 & $\frac{1}{3}$G(0,$ \, 1\sigma_{\mu}$)+$\frac{1}{3}$G(0,$ \, 2\sigma_{\mu}$)+$\frac{1}{3}$G(0,$ \, 3\sigma_{\mu}$)\\
      8 & $\frac{2}{3}$G(0,$ \, 1\sigma_{\mu}$)+$\frac{1}{3}$G(0,$ \, 3\sigma_{\mu}$)\\
      9 & $\frac{2}{3}$G(0,$ \, 2\sigma_{\mu}$)+$\frac{1}{3}$G(0,$ \, 3\sigma_{\mu}$)\\
      10 & G(0,$ \, 1\sigma_{\mu}$) [blue bands]\\
       & G(0,$ \, 2\sigma_{\mu}$) [red bands]\\
      11 & best of above (for each band)\\
      \hline
      \hline
    \end{tabular}
  \end{center}
\end{table}

\begin{figure*}
\includegraphics[width=\textwidth]{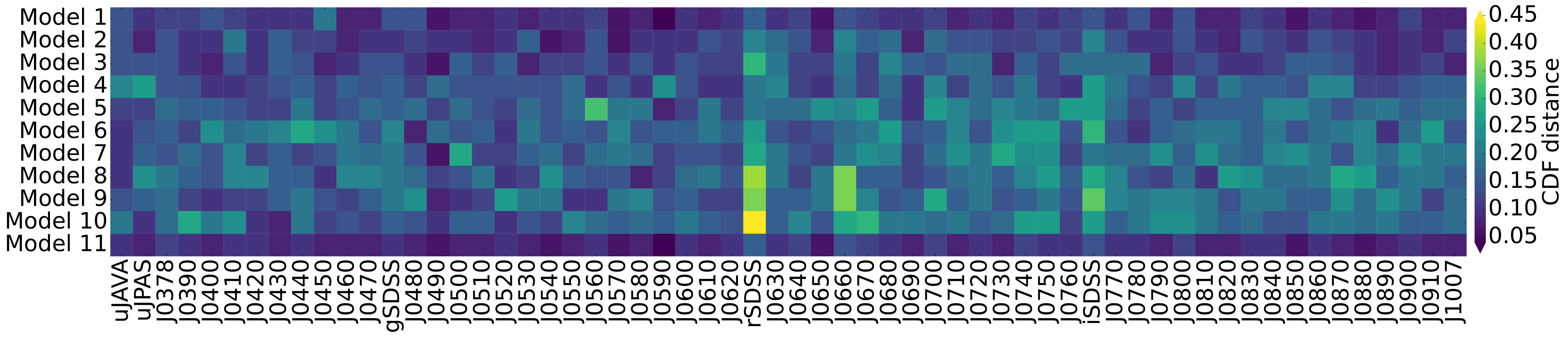}
\caption{Goodness of the fit for each noise model as a function of the passband. The colour code represents the maximum distance between the CDFs of the S/N of the miniJPAS-Superset stars, and the S/N of the synthetic photometry obtained using a given model, where lower values correspond to a better fit. Model 11 corresponds to the best-fitting model for each band.}
\label{fig:ks_noise_model}
\end{figure*}

To select the best noise model for each band, we use the sample of non-saturated miniJPAS-Superset stars, which are expected to have almost no variability, and compute the maximum differences between the cumulative distribution functions (CDF) of the S/N of the observations, and the S/N of the synthetic photometry obtained according to model \texttt{x}, with \texttt{x} ranging from 1 to 10. The best-fitting for each band corresponds then to the model that minimizes the difference between the CDFs, yielding model 11. In Fig. \ref{fig:ks_noise_model} we show the goodness of the fit (i.e. maximum distance between the CDFs) for each noise model as a function of the passband.

\begin{figure}
\includegraphics[width=\linewidth]{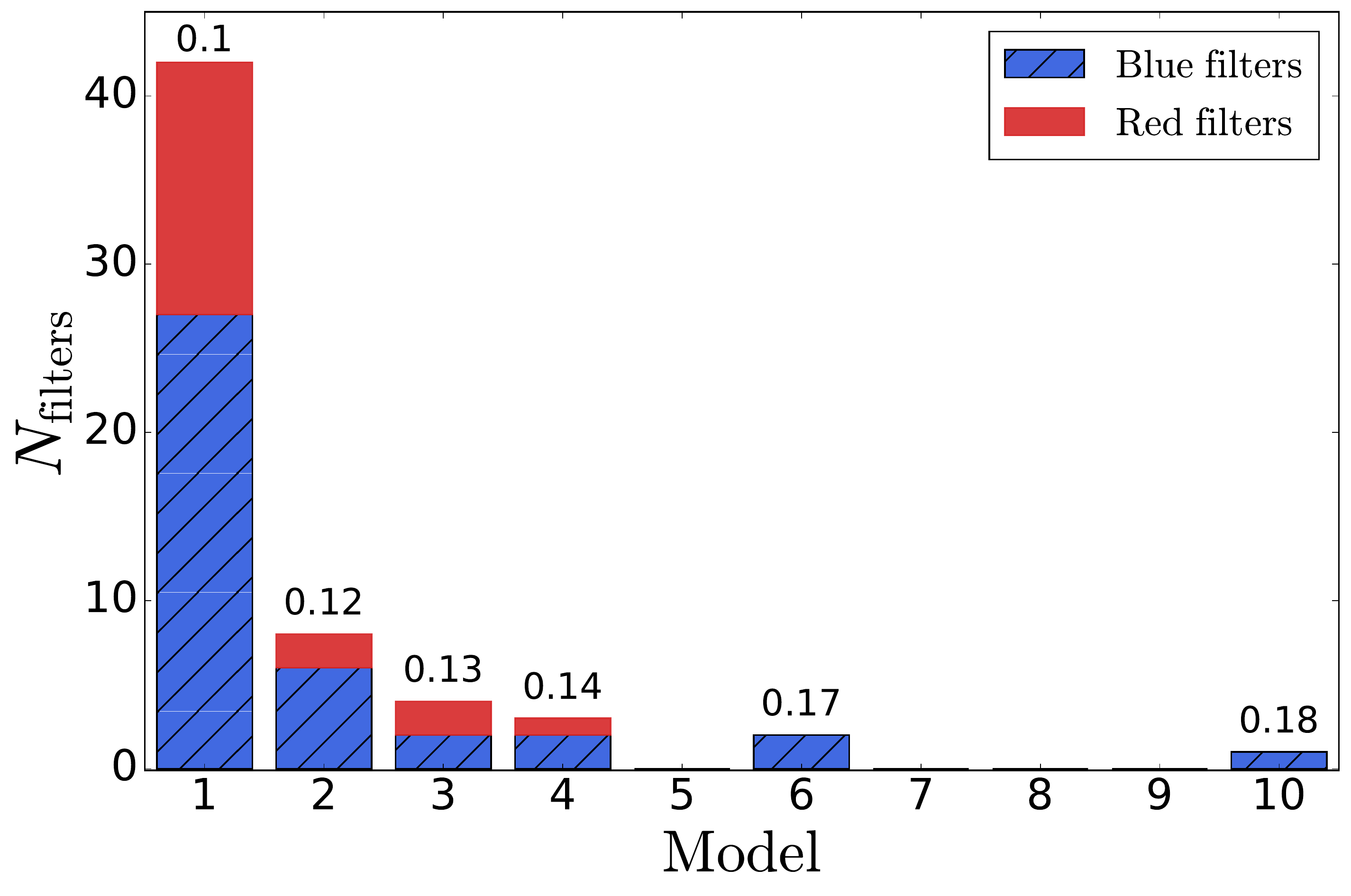}
\caption{Histogram of the number of filters that had their noise profiles best fitted by a given model. The frequencies are separated by the contributions from the blue and red bands. The number on top of the bars correspond to the median CDF distance over all bands for a given model.}
\label{fig:hist_noise_model}
\end{figure}

In Fig. \ref{fig:hist_noise_model} we provide the histogram of the number of filters which had their noise profile distributions best fitted by model \texttt{x}, with \texttt{x} ranging from 1 to 10. The contributions from the blue and red ($\lambda_{\mathrm{eff}}\geq7\,416$ {\AA}) bands are shown separately. The number on top of the bars corresponds to the median CDF distance over all bands for a given model; as a comparison, for model 11 this value is equal to 0.09. Although model 1 is able to reproduce the noise profile distributions of miniJPAS observations for most filters, models with larger widths are still preferable for some of the bands. These results agree with fig. A1 from \citet{Rosa21}, which shows that the noise profile distributions of the \texttt{PSFCOR} magnitudes for miniJPAS extended sources are globally well fitted by a Gaussian function with standard deviation equal to 1.4. They also report that some filters present a higher dispersion in the noise distribution, and that the miniJPAS errors are particularly underestimated in the red filters.

\begin{figure*}
\includegraphics[width=\textwidth]{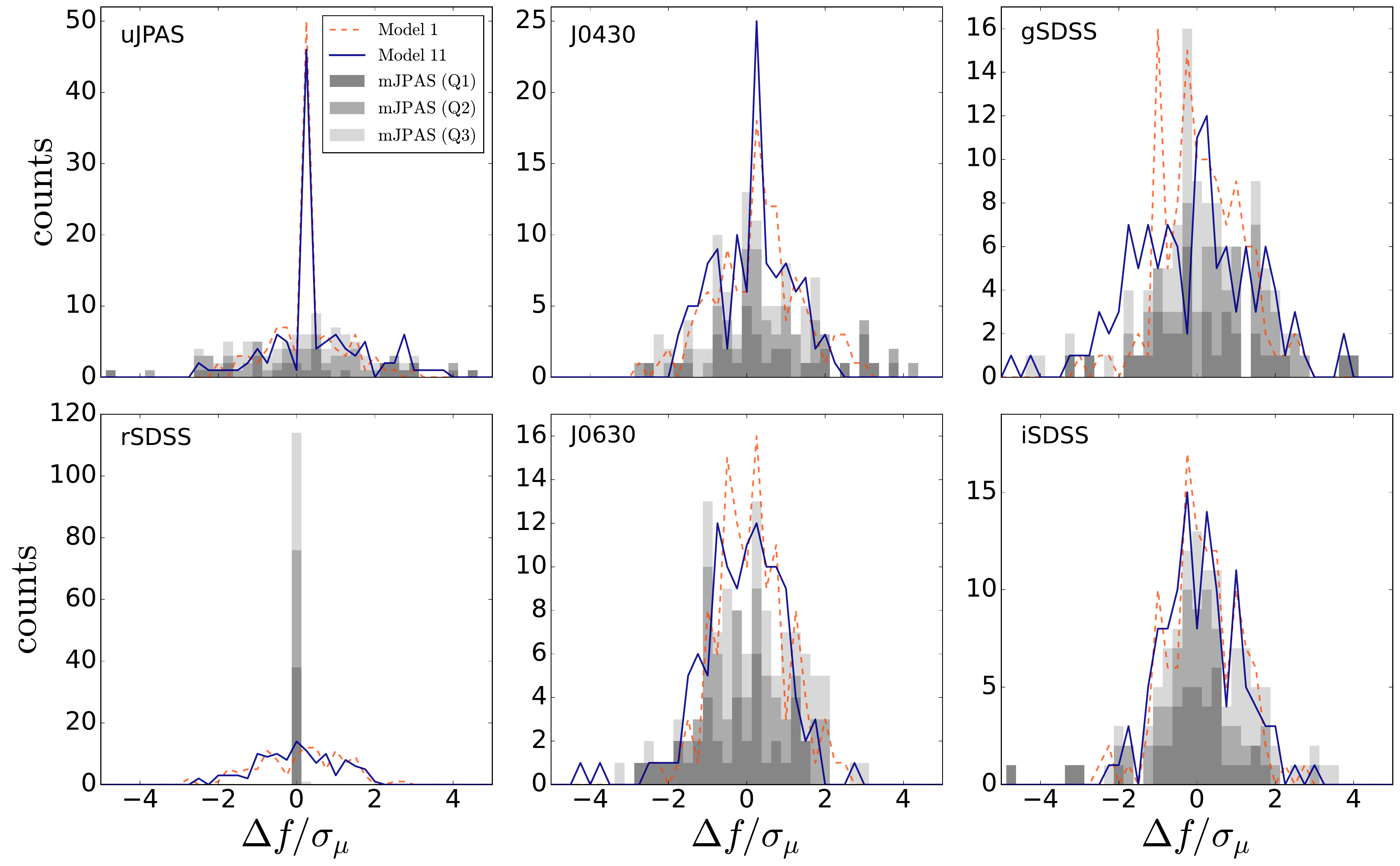}
\caption{Histograms of the differences between the observed and the synthetic fluxes ($\Delta f$) divided by the nominal errors ($\sigma_{\mu}$) for the miniJPAS-Superset stars. Here we show the distributions for six different bands: $uJPAS$, $J0430$, $gSDSS$, $rSDSS$, $J0630$ and $iSDSS$. The bars correspond to the miniJPAS point sources, whose magnitudes were divided into lower (Q1), median (Q2) and upper (Q3) tertiles, shown by the different shades of gray ranging from darker to lighter, respectively. We compare noise models 1 (orange dashed lines) and 11 (blue solid lines).}
\label{fig:hist_delta_flam_6bands}
\end{figure*}

In Fig. \ref{fig:hist_delta_flam_6bands} we show the histograms of the differences between the observed and the synthetic fluxes divided by the nominal errors of the spectroscopic stars at six different bands. We compare the synthetic fluxes generated based on models 1 and 11. In the case of the miniJPAS point sources, the synthetic fluxes are computed directly from Eq. \ref{eq:synt_flam} (i.e. without adding flux fluctuations), and the sources are divided in three equal parts according to their magnitudes. As we can see, there is no clear trend on the noise distribution of the brightest and faintest objects. The absence of scattering in the $r$-band for the miniJPAS observations is interpreted as due to a narrower error distribution, which reflects the fact that this band has a significantly higher $\left<S/N\right>$ than the narrow-bands, and was adopted as the reference band for the calibration of the miniJPAS images. 

The plots for the remaining 54 bands are shown in Appendix \ref{appx:hist_delta_flam}. Note that some of the bluest bands (e.g. $J0378$, $J0390$, $J0400$, and $J0410$ in Fig. \ref{fig:hist_delta_flam_1}) seem to have poorer fits when compared to the other bands. However, since the miniJPAs photometry is better calibrated than the Superset spectra, many of these differences actually come from calibration errors.

\subsection{ND patterns} \label{subsec:NO_patterns}

Following convention adopted in Sect. \ref{subsubsec:photo_sample}, non-detections originated from low signal-to-noise detections naturally appear in the mocks whenever $\mathrm{S/N}<1.25$ (case $(i)$); however, this is not so straightforward for the negative fluxes. 

If the noise fluctuations are sampled from very wide distributions, some of the resulting synthetic fluxes become negative, and are automatically flagged as NDs (case $(ii)$). However, these flagged bands alone are not able to properly reproduce the distribution of negative fluxes from the miniJPAS point sources. So we include in a consistent way more negative fluxes using the following prescription. For a given object from the mock catalog, we search in the non-sorted miniJPAS point-like sample for a source with similar $r$-band magnitude, and verify which of the bands have negative fluxes. Then, we apply this same pattern to the synthetic photometry but only to those bands which had already been flagged with low signal-to-noise ratios. This avoids an excessive degradation of valid detections in the mocks, which would typically affect more intensely fainter objects.

\subsection{Final mock catalogues} \label{subsec:Final_mocks}

The final mock catalogues are provided in two versions: fluxes per unit wavelength and magnitudes with the corresponding nominal errors, so as to reproduce a real catalogue of observations. Since one does not know \textit{a priori} the ``true'' distributions of objects in each region of the sky, to avoid any biases from our putative luminosity functions our final mocks for the classification of miniJPAS sources contain balanced samples of size 10k, 10k and 100k for the test, validation and training sets, respectively, and for each class of object.

\begin{figure*}
\includegraphics[width=\textwidth]{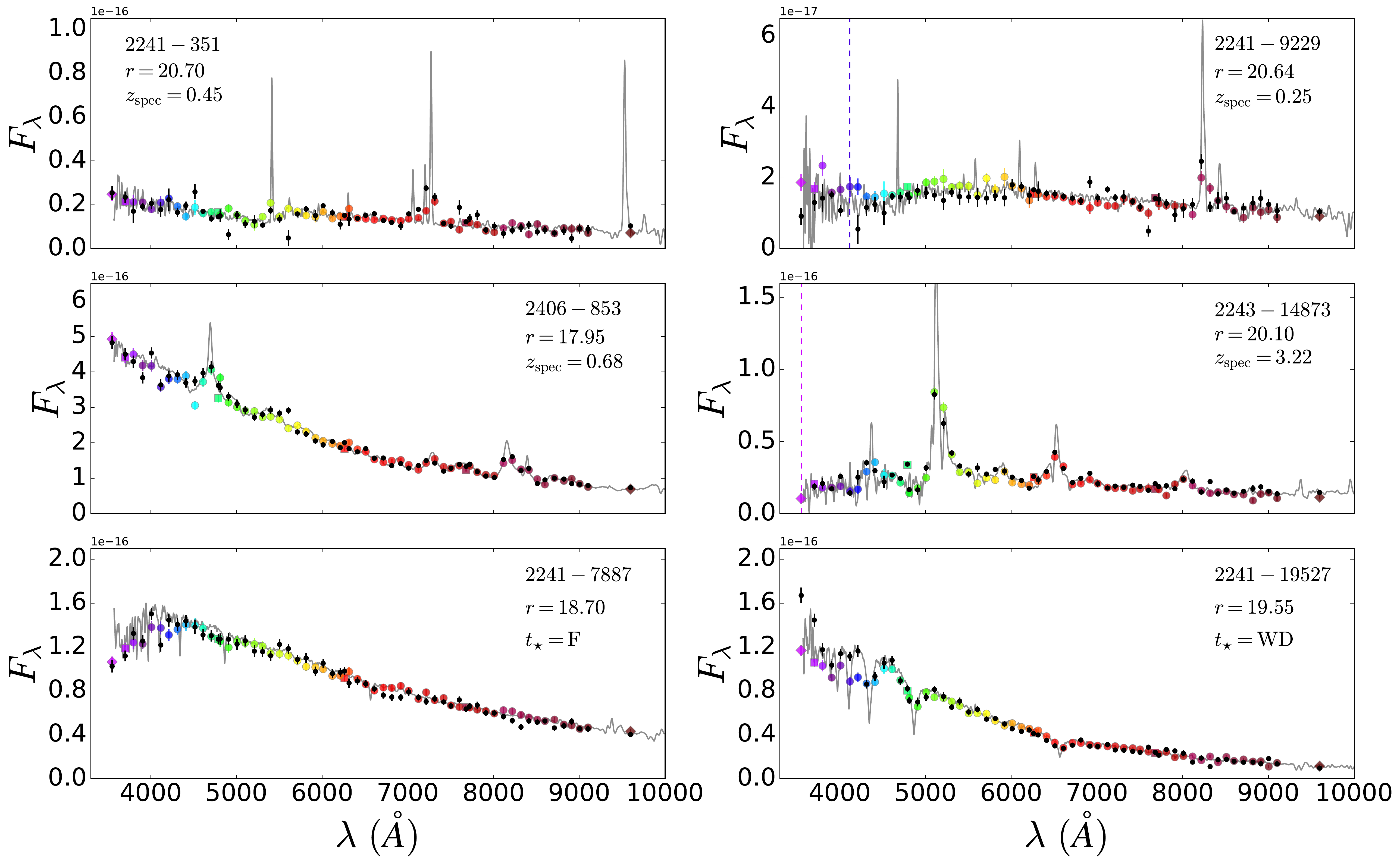}
\caption{Simulated photospectra of starburst galaxies (top), quasars (middle) and stars (bottom). The gray solid lines correspond to the smoothed SDSS spectra. Coloured diamonds, squares and dots correspond to the miniJPAS \texttt{APER3} fluxes in the medium, broad and narrow-bands, respectively. Black dots correspond to the synthetic fluxes generated using model 11 with their corresponding uncertainties. The miniJPAS objects are identified by their tile and number IDs; their $r$-band magnitudes, and spectroscopic redshifts $z_{\mathrm{spec}}$ (or stellar types $t_{\star}$) are also listed in the legend. The vertical dashed lines correspond to non-detections in the mocks. The fluxes are in units $\mathrm{erg} \, \mathrm{s}^{-1} \, \mathrm{cm}^{-2} \,
${\AA}$^{-1}$.}
\label{fig:flux_simu_composite}
\end{figure*}

In Fig. \ref{fig:flux_simu_composite} we provide some examples of synthetic photospectra generated with noise model 11 for galaxies, quasars and stars. As a comparison, we also show the corresponding SDSS spectra, and the miniJPAS \texttt{APER3} fluxes. The synthetic fluxes follow satisfactorily the miniJPAS observations, presenting some random statistical fluctuations within the expected levels, which is one of the key ingredients in our mocks.

The performances of the ML algorithms are validated on the test sets, as well as on the miniJPAS-Superset sample. The results of each classifier are also combined using a random forest algorithm (P\'erez-R\`afols et al. in prep. 2022b). Besides providing balanced test samples, we also generate a sample containing the relative incidence rates of objects per deg$^{2}$ to allow a more direct comparison with the performance of the classifiers on the miniJPAS spectroscopic sample.

\section{Mock validation} \label{sec:Results}

In this section we validate the mock catalogues by comparing the main properties of the synthetic fluxes with the observational features present in the miniJPAS point-like sample. Unless otherwise stated, these results correspond to the test sets generated using noise model 11.

\subsection{Magnitude limits} \label{subsec_mag_lim}

\begin{figure}
\includegraphics[width=\linewidth]{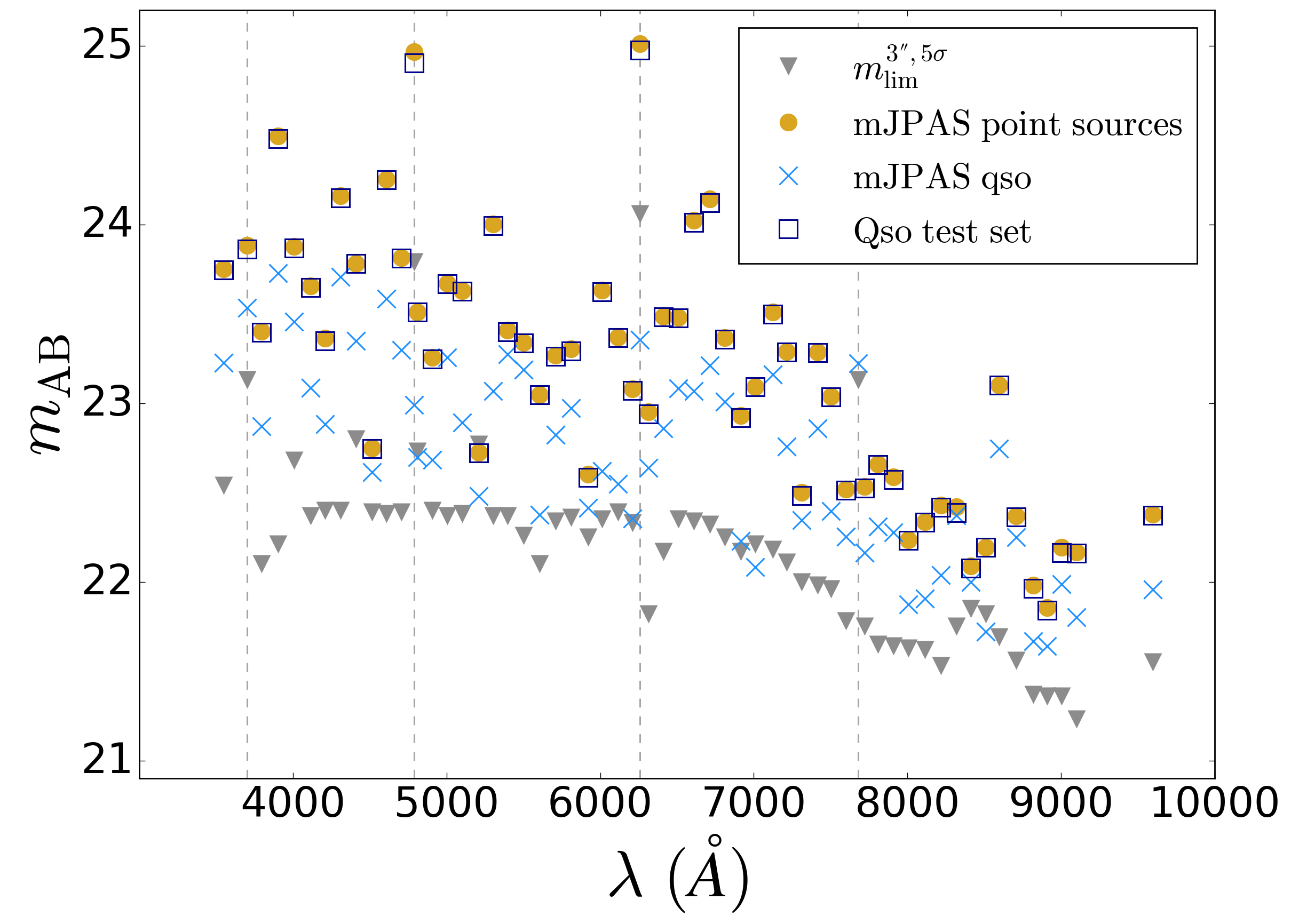}
\caption{Maximum magnitudes reached by each filter in the quasar test set (dark blue open squares). The yellow dots and light blue crosses represent the depths reached by the miniJPAS point-like sample (over all tiles) and the miniJPAS-Superset quasars, respectively. As a comparison, we also show the J-PAS targeted minimum depth within an aperture of 3$\arcsec$ (gray down-pointing triangles). The vertical gray dashed lines indicate the positions of the four broad-bands ($uJPAS$, $gSDSS$, $rSDSS$ and $iSDSS$, respectively).}
\label{fig:qso_test_mag_max}
\end{figure}

The magnitude limits reached by each band in the quasar test set are shown in Fig. \ref{fig:qso_test_mag_max}. We compare these depths with the maximum magnitudes reached by the miniJPAS point-like sample (over all tiles) and the miniJPAS-Superset quasars. Note that these magnitude limits are obtained by considering only valid detections (i.e. $\mathrm{S/N}>1.25$ and $F_{\lambda, \, \mu}^{synt}>0$). The depth reached in the quasar test set is in great accordance with the miniJPAS observations. As a comparison, we also show the J-PAS theoretical minimum depth (considering $\mathrm{S/N}=5$ within an aperture of 3$\arcsec$).

Although not shown here, the magnitude limits reached in the training, validation and 1-deg$^{2}$ sets are very similar to what we have demonstrated for the test set -- apart, of course, from some fluctuations, as already expected due to the different sample size and different realizations of the luminosity function in each set. In the case of galaxies, the mocks are able to reach the same depths as the miniJPAS point sources in all bands. On the other hand, the mocks of stars are typically brighter for most bands; in particular, the $g$ and $r$-bands do not reach the same depths as the point sources. We attribute this effect due to the shape of the luminosity function, which predicts a significant decrease in the number of stars at $r>21.5$ (as shown in Fig. \ref{fig:lf_star}).


\subsection{Signal-to-noise ratio} \label{subsec_snr}

\begin{figure}
\includegraphics[width=\linewidth]{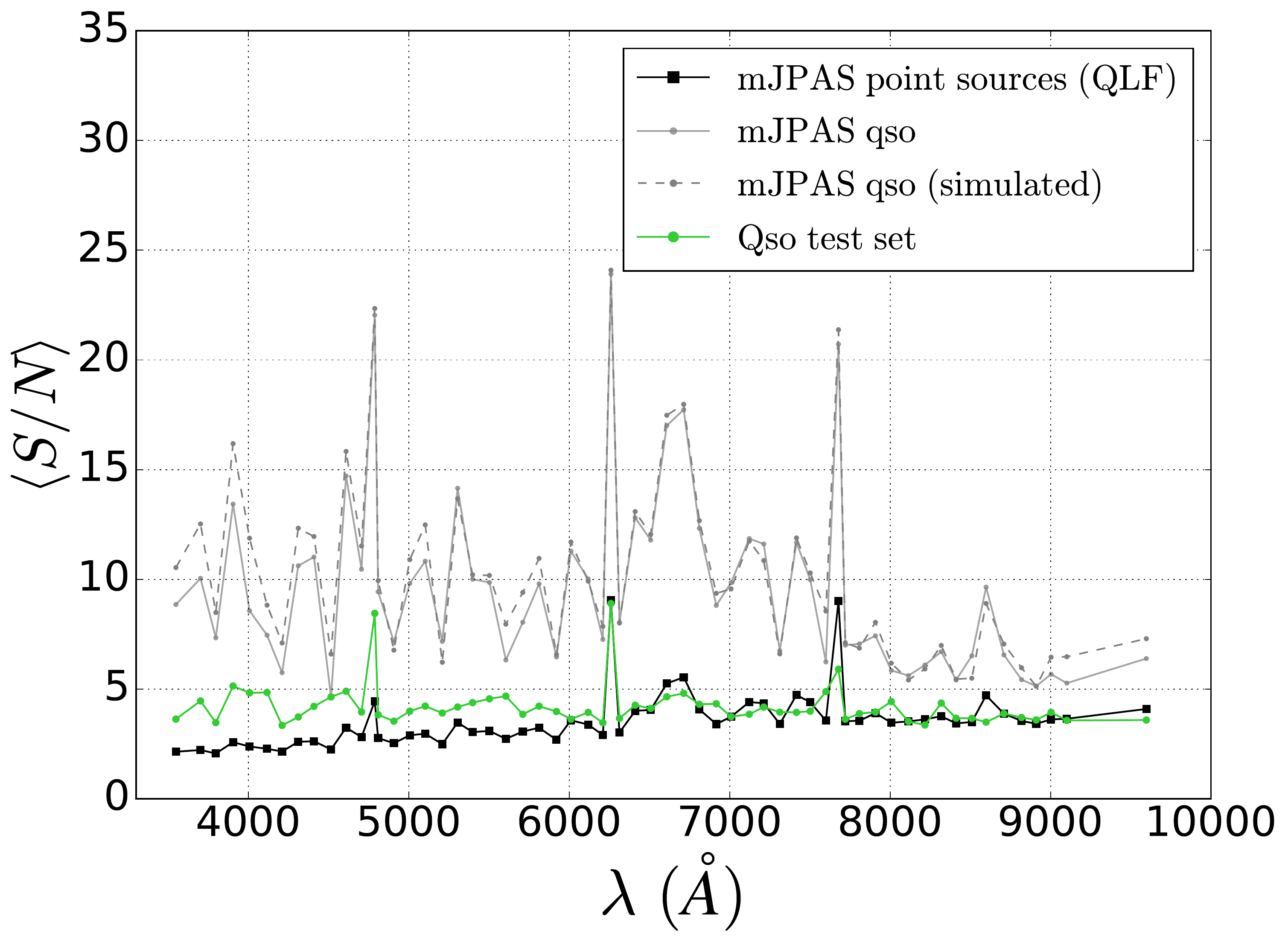}
\caption{Median signal-to-noise ratio as a function of the filter for the subsample of miniJPAS point sources randomly selected according to the QLF (black line), miniJPAS-Superset quasars (gray solid line) and quasars from the test set (green solid line). As a comparison, we also show the median signal-to-noise ratio obtained for the synthetic fluxes of the miniJPAS-Superset quasars (gray dashed line).}
\label{fig:qso_test_snr}
\end{figure}

In Fig. \ref{fig:qso_test_snr} we compare the median S/N distributions per band for the miniJPAS point sources randomly selected according to the QLF, miniJPAS-Superset quasars, and quasars from the test set. As already expected, miniJPAS-Superset quasars have typically larger S/N distributions than the mocks. On one hand, we can attribute this to the lack of a representative sample of faint objects in the Superset sample, which increases $\left<S/N \right>$. Equivalently, the LF yields a fairer sample of faint sources in the mocks (with typically larger associated errors) which ends up dominating (and, subsequently, lowering down) the median S/N distribution. As a consequence, the signal-to-noise ratio in the test set becomes more representative of the miniJPAS point sources. 

We also identify similar modulations in the S/N of adjacent filters for both the observations and simulations. Since the presence of these modulations are directly associated to the miniJPAS survey strategy (e.g. filter exposure times, net effect of sky brightness and final number of combined images), having them in the mocks indicates that we have successfully matched the S/N distributions from the observations into the mocks. This can be better seen by comparing the median S/N achieved by the observations and the synthetic fluxes of the miniJPAS-Superset quasars.

\begin{figure}
\includegraphics[width=\linewidth]{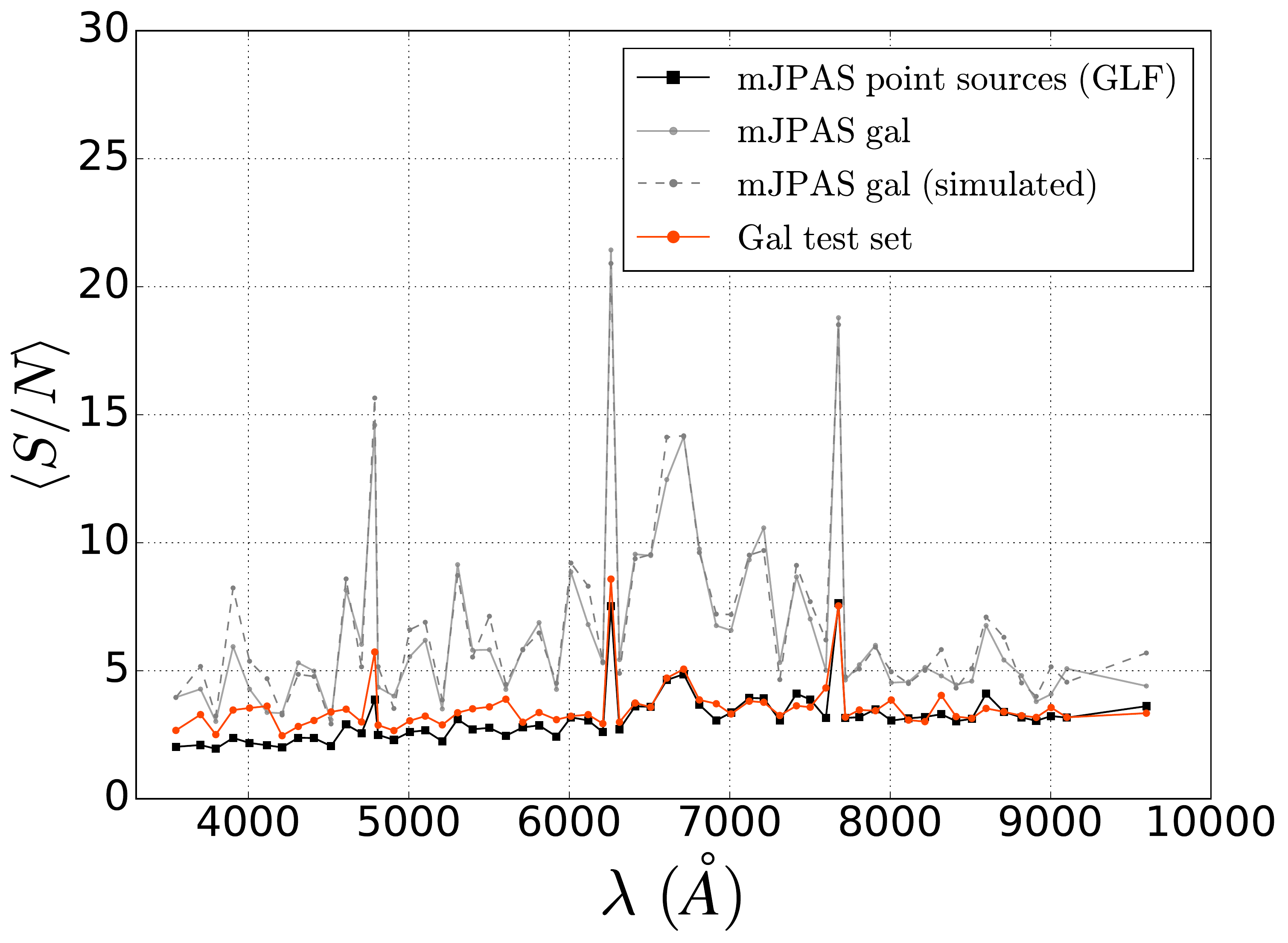}
\caption{Median signal-to-noise ratio as a function of the filter for the subsample of miniJPAS point sources selected according to the GLF (black line), miniJPAS-Superset galaxies (gray solid line) and galaxies from the test set (orange solid line). As a comparison, we also show the median signal-to-noise ratio obtained for the synthetic fluxes of the miniJPAS-Superset galaxies (gray dashed line).}
\label{fig:gal_test_snr}
\end{figure}

\begin{figure}
\includegraphics[width=\linewidth]{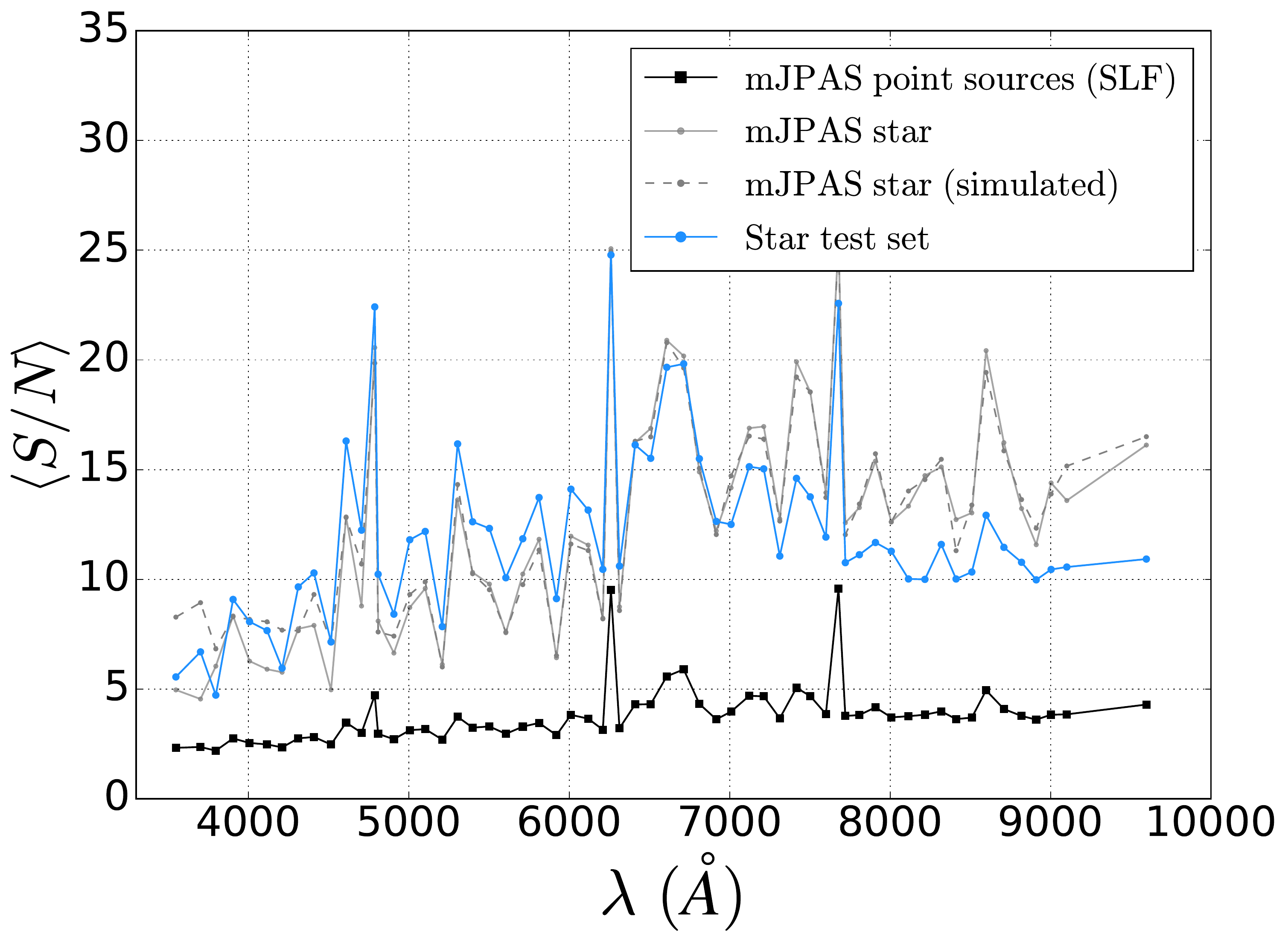}
\caption{Median signal-to-noise ratio as a function of the filter for subsample of miniJPAS point sources selected according to the SLF (black line), miniJPAS-Superset stars (gray solid line) and stars from the test set (light blue solid line). As a comparison, we also show the median signal-to-noise ratio obtained for the synthetic fluxes of the miniJPAS-Superset stars (gray dashed line).}
\label{fig:star_test_snr}
\end{figure}

In Fig. \ref{fig:gal_test_snr} and \ref{fig:star_test_snr} we show the median S/N distributions for the galaxy and star test sets, respectively. Again, we observe the same patterns of modulations both in the observations and in the mocks. Note, however, that in the case of stars the median S/N distribution for the test set follows the spectroscopic sample more closely than the miniJPAS point sample. This effect is attributed to the decrease in the number counts predicted by the SLF for $r\gtrsim21.5$ (see Fig. \ref{fig:lf_star}), which results in a typically brighter sample in comparison to the simulated samples of quasars and galaxies -- whose luminosity functions tend to predict more objects in the faint end. 

\begin{figure*}
\includegraphics[width=\linewidth]{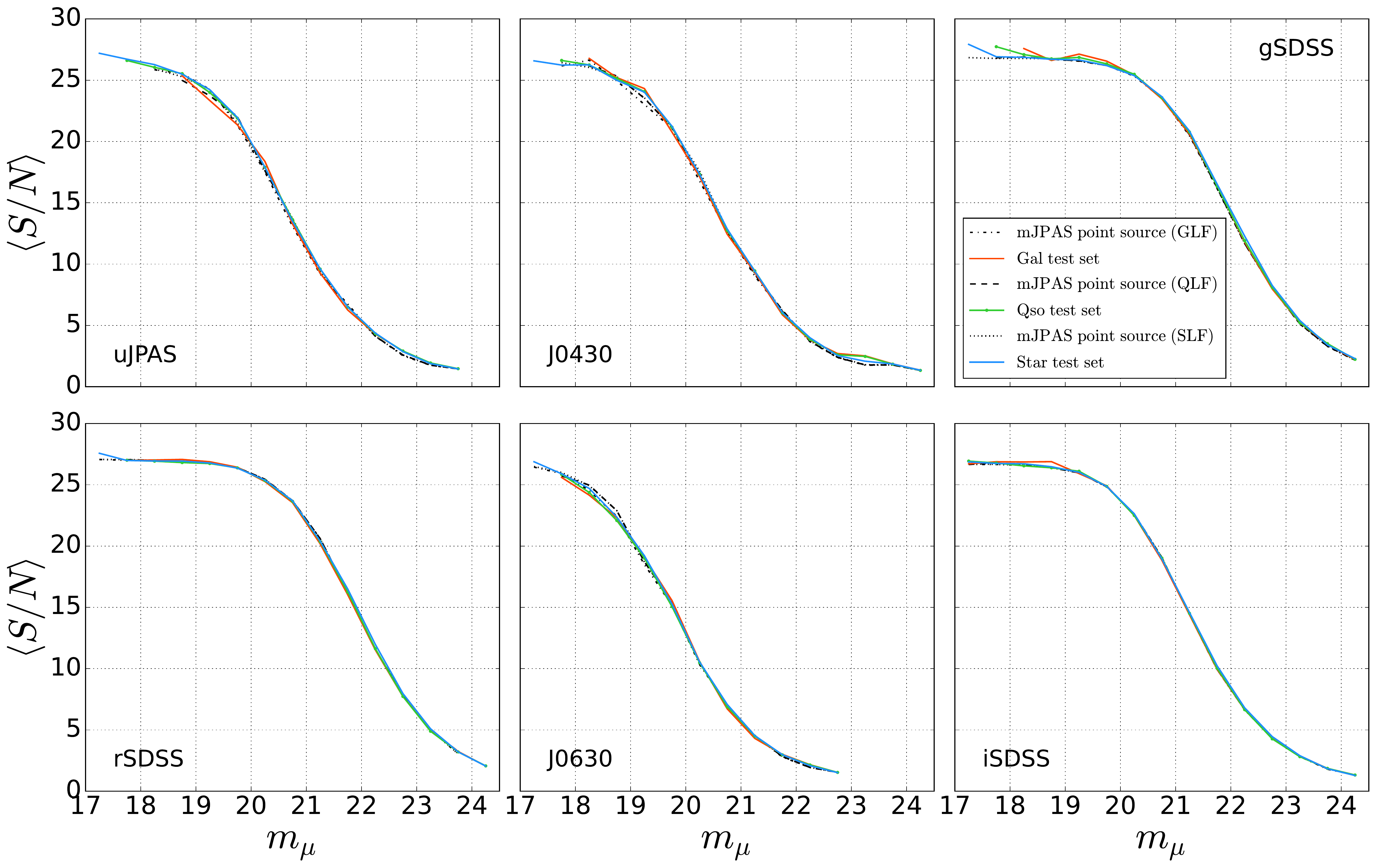}
\caption{Median signal-to-noise ratio as a function of the magnitude at six different bands ($uJPAS$, $J0430$, $gSDSS$, $rSDSS$, $J0630$ and $iSDSS$) for galaxies (orange solid line), quasars (green solid line) and stars (light blue solid line). As a comparison, we show the subsamples of miniJPAS point sources that were randomly selected so as to reproduce equivalent luminosity distributions for each type of source.}
\label{fig:snr_mag_test}
\end{figure*}

In Fig.\ref{fig:snr_mag_test} we evaluate how $\left<S/N\right>$ varies as a function of the magnitude in the test set for six different bands ($uJPAS$, $J0430$, $gSDSS$, $rSDSS$, $J0630$ and $iSDSS$). As a comparison, we also show the subsamples of miniJPAS point sources that were randomly selected according to a given luminosity function. Results are provided for galaxies, quasars and stars. All bands shown here present the same general trend of having a decreasing $\left<S/N\right>$ for fainter magnitudes. This demonstrates again that the mocks yield comparable signal-to-noise ratio properties as verified in a miniJPAS point-like subsample with equivalent luminosity distributions.

\subsection{Non-detections} \label{subsec_NO_ND}

\begin{figure}
\includegraphics[width=\linewidth]{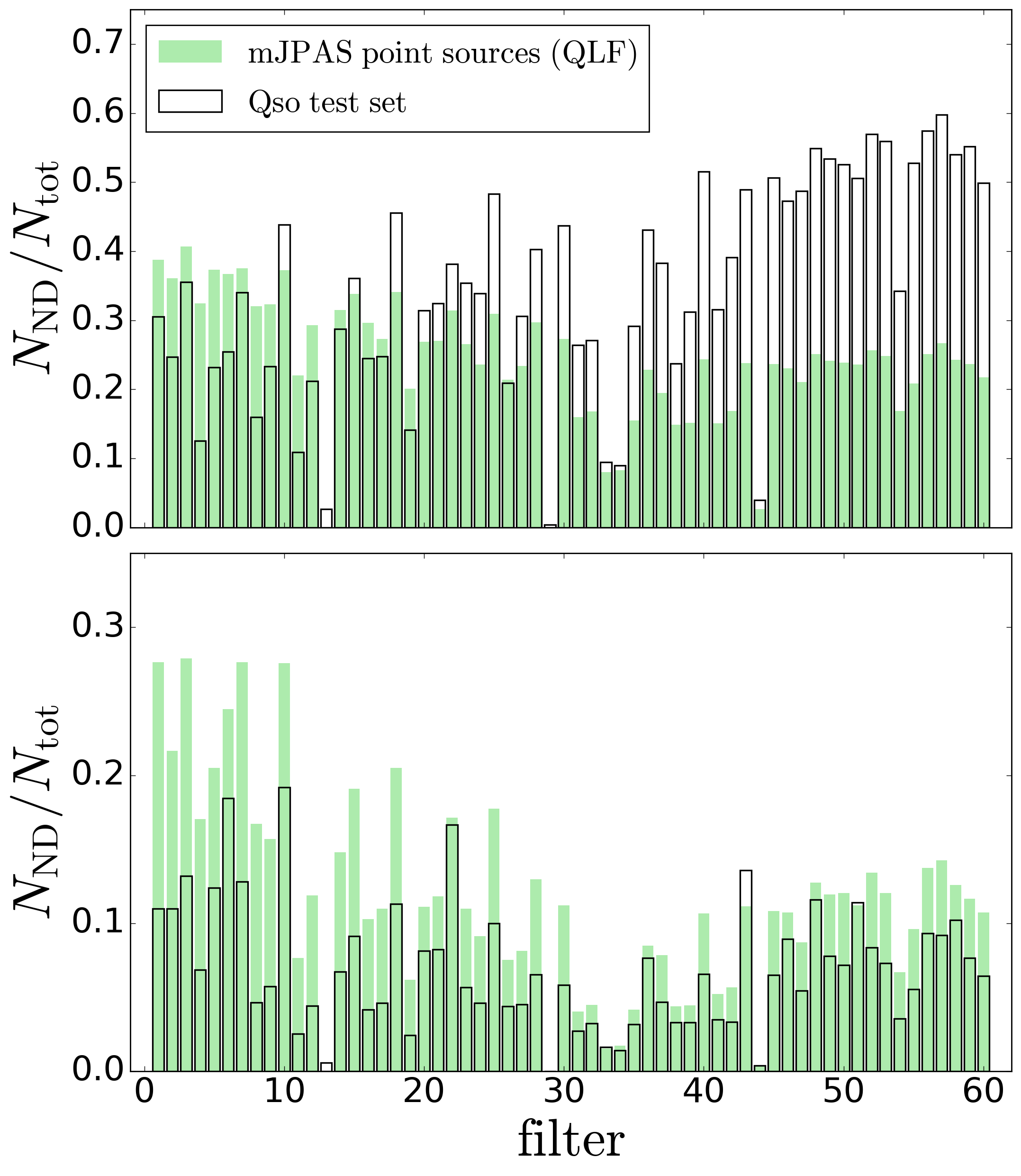}
\caption{Non-detection fraction per band for the quasar test set (black lines), and the subsample of point sources randomly selected according to the QLF (green solid bars). The x-axis represents the filter indices, which are ordered according to the effective wavelength. We separate the contributions from low signal-to-noise ratio (upper panel) and negative (lower panel) fluxes.}
\label{fig:qso_test_no_nd}
\end{figure}

In Fig. \ref{fig:qso_test_no_nd} we compare the number density of non-detections per band for the quasar test set and the subsample of miniJPAS point sources randomly selected according to the QLF. We separate the contributions from low S/N sources and negative fluxes. 


The pattern of non-detections is in general well reproduced in the quasar test set; in particular, we do not seem to include a too large fraction of negative fluxes. Nevertheless, the reddest bands, which correspond to filter indices $>40$, seem to be affected by a large fraction of more noisy objects than the observations. These noisier quasars dominate the medium signal-to-noise ratios at magnitudes $r>22.0$. Similarly to what we found for quasars, the galaxy test set presents a large fraction of more noisy objects than the observations, but the fraction of negative fluxes is not overestimated. On the other hand, the mocks of stars result in larger fractions of objects with higher signal-to-noise ratios than the observations, and equivalently smaller fractions of negative fluxes.

\begin{figure}
\includegraphics[width=\linewidth]{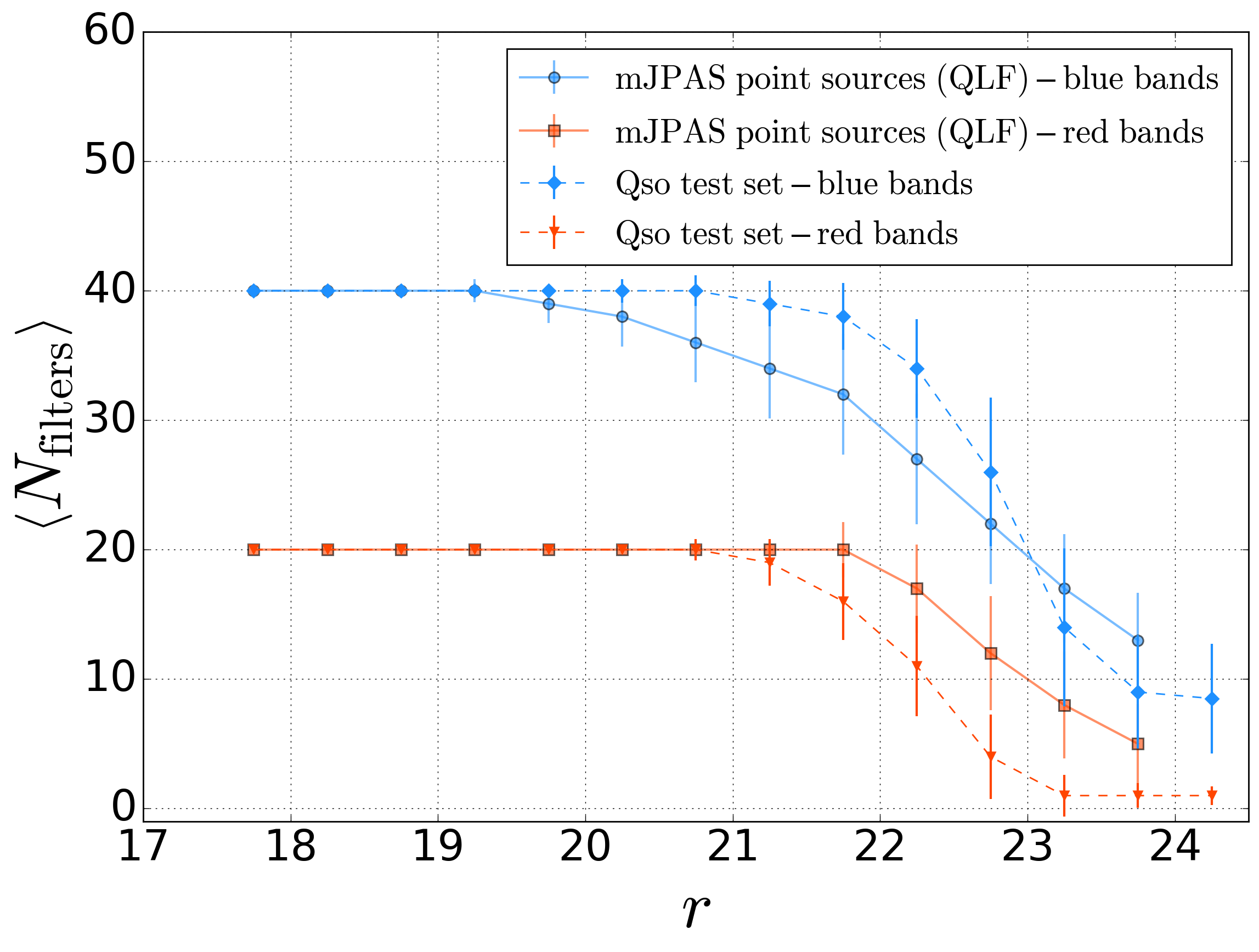}
\caption{Median number of filters with valid detections as a function of $r$-band magnitude. We compare the number of detections present in the miniJPAS point sources randomly selected according to the QLF (solid lines) with that present in the quasar test set (dashed lines). The error bars correspond to the standard deviations. We divide the contributions between blue and red ($\lambda_{\mathrm{eff}}\geq 7\,416$ {\AA}) bands.}
\label{fig:Nfilt_qso_test}
\end{figure}

An illustration of how the number of observed filters is degraded for fainter $r$-band magnitudes is provided in Fig. \ref{fig:Nfilt_qso_test}. As a comparison, we show the median number of filters with valid detections per magnitude bin for the quasar test set and the subsample of miniJPAS point sources randomly selected according to the QLF. We also divide the contributions between blue and red ($\lambda_{\mathrm{eff}}\geq 7\,416$ {\AA}) bands. When compared with the miniJPAS point sources, the simulated fluxes in the blue bands present smaller fractions of non-detections in the range $20\leq r \leq 22.5$, while presenting a strict suppression of detections in the reddest bands at the faint end ($r>21.5$).

\subsection{Quasar offsets} \label{subsec:qso_variability}

Finally, since quasars have long been known as intrinsically variable sources, one question that naturally arises is whether the miniJPAS quasars present any evidence of variability. 

\begin{figure*}
\includegraphics[width=\linewidth]{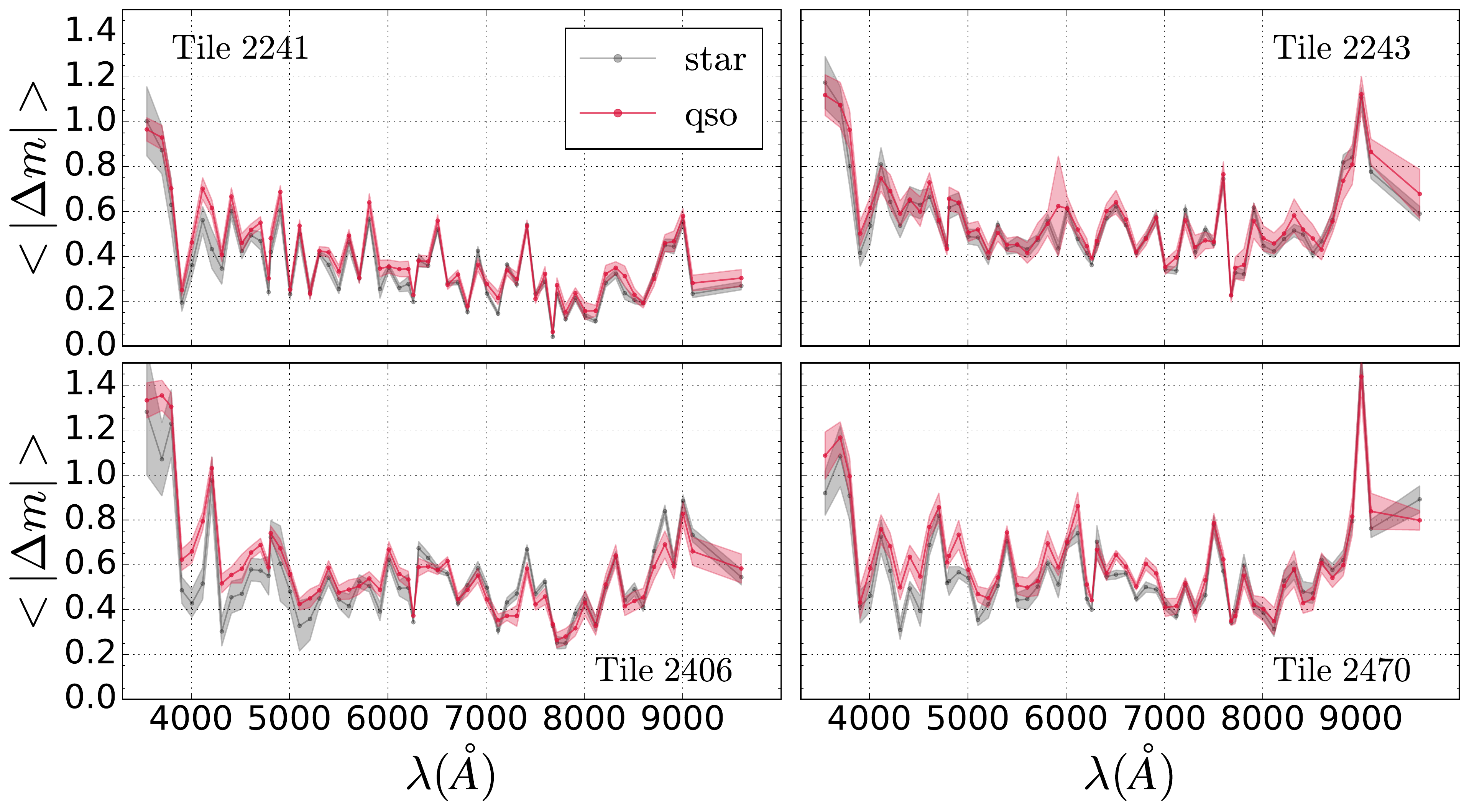}
\caption{Medium absolute differences between miniJPAS magnitudes and SDSS DR16Q synthetic magnitudes (without noise fluctuations) for quasars (red) and stars (gray) as a function of the tile. The solid regions correspond to the standard deviations.}
\label{fig:qso_variability}
\end{figure*}

To investigate this hypothesis, we employ SDSS DR16Q (\citealt{Myers15}; \citealt{Lyke20}) spectra, which constitute an improved spectrophotometric calibration for BOSS quasars (see e.g. \citet{Margala16} for further details). We then compute the medium absolute differences (and corresponding standard deviations) of the miniJPAS magnitudes and synthetic magnitudes (without noise fluctuations) for Superset quasars, and compare this offsets with the ones obtained for stars. These differences are computed as a function of the tile, and the results are shown in Fig. \ref{fig:qso_variability}. In essence, any evidence for quasar variability would be indicated by variations larger than the error bars of stars. However, we found no significant evidence of quasar variability. Hopefully, future observations will allow us to make a better assessment of  quasar variability in the AEGIS field.

\section{Discussion} \label{sec:Discussion}

Training machine learning algorithms in the presence of simulated fluxes can be more advantageous than a training based solely on real data sets (composed of e.g. spectroscopically confirmed sources), because the latter might include spectral misclassifications, as well as a non-fair distribution of redshifts and luminosities that can bias the resulting ML models, and consequently bias the classification. In particular, since the number of spectroscopically confirmed sources in the miniJPAS area surveyed by miniJPAS is not sufficiently large nor complete, the mock catalogues described in this work are crucial to perform the selection of the miniJPAS quasar candidates.

The results shown in this paper suggest that the mocks have successfully reached similar depths and levels of signal-to-noise ratio as the miniJPAS point sources. To further to optimize the methods developed for generating mock catalogues, we highlight in the following some future improvements:

\begin{enumerate}
    \item expand the library of spectra to include other types of sources that may be misrepresented in the current version. For instance, we lack a fair sample of some stellar spectral types (such as white dwarfs), galaxies brighter than $r=18.7$ and at redshifts larger than $z=0.9$, Lyman-break galaxies and Lyman-$\alpha$ emitters, as well as type-II AGNs and red quasars. The inclusion of these objects in the training sets will allow us to refine the machine learning classifiers, and better assess the contaminants within the sample of quasar candidates;
    \item better assess the luminosity priors for galaxies and stars. In particular, model the distributions of e.g. blue and red galaxies in more detail;
    \item impose less conservative selection criteria in the miniJPAS catalog, allowing, for instance, sources with some sort of flag, and explore different cuts in stellarity to assess how the performance of the classifiers changes; 
    \item include observations from other wavelengths (when available) -- such as infrared information from the Wide-field Infrared Survey Explorer (WISE; \citealt{Wright10}) and proper motions from Gaia (\citealt{Gaia16}), to assist the classification. The mocks could also be supplemented with morphological parameters (coming from the modeling of miniJPAS sources) to improve e.g. the separation between unresolved galaxies and quasars. Nevertheless, combining all this information with the optical spectra poses an interesting challenge.
\end{enumerate}

Finally, observations with the WEAVE-QSO survey (\citealt{Pieri}) will provide us with spectra of high-redshift quasars and Lyman-$\alpha$ systems, with unprecedented spectral resolution (mostly R=5\,000 but also R=20\,000), allowing us to improve the simulated fluxes (particularly at the faint end). The WEAVE-QSO spectroscopic follow-ups will enlighten us on the most critical improvements to the mock catalogues, as we will be able to better assess the performance of the ML classifiers, and confirm (or exclude) sources identified as potential quasar candidates. Moreover, J-PAS will soon start gathering data, which will further help us improve the noise models.

\subsection{Additional applications} \label{subsec:further_applications}

Our mock catalogues are also relevant for further interesting applications. For instance, in Queiroz et al. (in prep.) we present a novel technique to estimate the photometric redshifts (photo-zs) of the miniJPAS quasar candidates based on a best-fitting model for the quasar photospectrum in terms of the so-called eigenspectra derived from a principal component analysis. Since we do not have a large enough sample of spectroscopic confirmed miniJPAS quasars, the mocks are essential to validate the performance of this photo-z method.

Moreover, early operation of the J-PAS survey may be conducted with less than 60 filters. This means that, in addition to the classification that we perform for miniJPAS point sources, the mocks will be fundamental to forecast the accuracy with which we will be able to find quasars with J-PAS, even in the case where we have observations in less than 60 bands. Although applied to the J-PAS photometric system, the methodology presented here can be easily adapted to build mocks for other (narrow-to-medium-band) photometric systems.

\section{Conclusions} \label{sec:Conclusions}

This paper is part of a series of manuscripts that aim at developing tools to classify miniJPAS sources in preparation for J-PAS, and identify quasar candidates. Since no real data set exists at present that is sufficiently large or complete to serve that purpose, constructing mock catalogues is crucial to properly train our machine learning classifiers, and assess their performances until such a time when both our photometric data and spectroscopic follow-up data reach sufficient size.

In this paper we present the pipeline to generate simulated photospectra of quasars, galaxies and stars containing the same signal-to-noise ratio distributions expected for miniJPAS point-like sources. Starting from synthetic fluxes obtained by the convolution of SDSS spectra with the J-PAS photometric system, we show how to incorporate realistic observational features by $(i)$ imposing that the relative incidence rates of the different classes of objects in the mocks follow the expected count numbers from putative luminosity functions; $(ii)$ carefully modeling the noises in all bands, and adding compatible levels of noise to the synthetic photometry; and $(iii)$ adding the patterns of non-detections (which dominate the faint end). Our results indicate that the miniJPAS fluxes in each band are best described by different noise profile distributions, but typically 1 to 1.5-$\sigma$ Gaussian functions can properly fit the uncertainties in most filters. 

Our final mock catalogues demonstrated the capability of correctly reaching the expected depths in all bands, and matching the signal-to-noise ratio distributions from the observations. These mock catalogues are invaluable for many scientific applications within the J-PAS collaboration, and will also be important for the whole astronomical community, as the procedure outlined here can be easily adapted to serve the purposes of other photometric surveys.

Once J-PAS is fully operational, we will be able to train the machine learning classifiers using directly real data. However, the mocks will not lose their crucial aspect of helping in the refinement of our methods, and avoiding misleading conclusions when analysing real observations. Even then, the mocks will perdure as our allies in the process of unraveling the cosmos in the search for quasars.

\section*{Acknowledgements}

This paper has gone through internal review by the J-PAS collaboration. C.Q. acknowledges financial support from the Brazilian funding agencies FAPESP (grants 2015/11442-0 and 2019/06766-1) and Coordena\c{c}\~ao de Aperfei\c{c}oamento de Pessoal de N\'ivel Superior (Capes) -- Finance Code 001. I.P.R., M.P.P. and S.S.M. were supported by the Programme National de Cosmologie et Galaxies (PNCG) of CNRS/INSU with INP and IN2P3, co-funded by CEA and CNES, the A*MIDEX project (ANR-11-IDEX-0001-02) funded by the ``Investissements d'Avenir'' French Government program, managed by the French National Research Agency (ANR), and by ANR under contract ANR-14-ACHN-0021. G.M.S., R.M.G.D. and L.A.D.G. acknowledge support from the State Agency for Research of the Spanish MCIU through the ``Center of Excellence Severo Ochoa'' award to the Instituto de Astrof\'isica de Andaluc\'ia (SEV-2017-0709) and the project PID2019-109067-GB100. J.C.M. and S.B. acknowledge financial support from Spanish Ministry of Science, Innovation, and Universities through the project PGC2018-097585-B-C22. A.F.S. acknowledges support from the Spanish Ministerio de Ciencia e Innovaci\'on through project PID2019-109592GB-I00 and the Generalitat Valenciana project PROMETEO/2020/085. R.A.D. acknowledges partial support support from CNPq grant 308105/2018-4. AE acknowledges the financial support from the Spanish Ministry of Science and Innovation and the European Union - NextGenerationEU through the Recovery and Resilience Facility project ICTS-MRR-2021-03-CEFCA. LSJ acknowledges support from CNPq (304819/2017-4) and FAPESP (2019/10923-5). J.V. acknowledges the technical members of the UPAD for their invaluable work: Juan Castillo, Tamara Civera, Javier Hern\'andez, \'Angel L\'opez, Alberto Moreno, and David Muniesa. 

Based on observations made with the JST250 telescope and PathFinder camera for the miniJPAS project at the Observatorio Astrof\'{\i}sico de Javalambre (OAJ), in Teruel, owned, managed, and operated by the Centro de Estudios de F\'{\i}sica del  Cosmos de Arag\'on (CEFCA). We acknowledge the OAJ Data Processing and Archiving Unit (UPAD) for reducing and calibrating the OAJ data used in this work. Funding for OAJ, UPAD, and CEFCA has been provided by the Governments of Spain and Arag\'on through the Fondo de Inversiones de Teruel; the Aragonese Government through the Research Groups E96, E103, E16\_17R, and E16\_20R; the Spanish Ministry of Science, Innovation and Universities (MCIU/AEI/FEDER, UE) with grant PGC2018-097585-B-C21; the Spanish Ministry of Economy and Competitiveness (MINECO/FEDER, UE) under AYA2015-66211-C2-1-P, AYA2015-66211-C2-2, AYA2012-30789, and ICTS-2009-14; and European FEDER funding (FCDD10-4E-867, FCDD13-4E-2685). Funding for the J-PAS Project has also been provided by the Brazilian agencies FINEP, FAPESP, FAPERJ and by the National Observatory of Brazil with additional funding provided by the Tartu Observatory and by the J-PAS Chinese Astronomical Consortium. Funding for the Sloan Digital Sky Survey III/IV has been provided by the Alfred P. Sloan Foundation, the U.S. Department of Energy Office of Science, and the Participating Institutions. SDSS-III/IV acknowledge support and resources from the Center for High Performance Computing  at the University of Utah. The SDSS website is \url{www.sdss.org}. SDSS is managed by the Astrophysical Research Consortium for the Participating Institutions of the SDSS Collaboration including the Brazilian Participation Group, the Carnegie Institution for Science, Carnegie Mellon University, Center for Astrophysics | Harvard \& Smithsonian, the Chilean Participation Group, the French Participation Group, Instituto de Astrof\'isica de Canarias, The Johns Hopkins University, Kavli Institute for the Physics and Mathematics of the Universe (IPMU) / University of Tokyo, the Korean Participation Group, Lawrence Berkeley National Laboratory, Leibniz Institut f\"ur Astrophysik Potsdam (AIP), Max-Planck-Institut f\"ur Astronomie (MPIA Heidelberg), Max-Planck-Institut f\"ur Astrophysik (MPA Garching), Max-Planck-Institut f\"ur Extraterrestrische Physik (MPE), National Astronomical Observatories of China, New Mexico State University, New York University, University of Notre Dame, Observat\'ario Nacional / MCTI, The Ohio State University, Pennsylvania State University, Shanghai Astronomical Observatory, United Kingdom Participation Group, Universidad Nacional Aut\'onoma de M\'exico, University of Arizona, University of Colorado Boulder, University of Oxford, University of Portsmouth, University of Utah, University of Virginia, University of Washington, University of Wisconsin, Vanderbilt University, and Yale University. 

The authors would like to thank Alvaro Alvarez-Candal, Juan Antonio Fern\'andez Ontiveros, and Mirjana Povic for useful suggestions and comments, and the feedbak of Joel Bregman.

This research made use of the following python packages: \texttt{Astropy} (\citealt{Astropy13}, \citeyear{Astropy18}), \texttt{Matplolib} (\citealt{Matplotlib}), \texttt{NumPy} (\citealt{NumPy}) and \texttt{SciPy} (\citealt{SciPy}).

\section*{Data availability}

We will make our mock catalogues publicly available upon publication.




\bibliographystyle{mnras}
\bibliography{biblio} 



\section*{Affiliations}
\noindent
\textit{
$^{12}$ Shanghai Astronomical Observatory, Chinese Academy of Sciences, 80 Nandan Rd., Shanghai 200030, China\\
$^{13}$ Instituto de F\'isica de Cantabria (CSIC-UC), Avda. Los Castros s/n. 39005, Santander, Spain\\
$^{14}$ Unidad Asociada ``Grupo de Astrof\'isica Extragal\'actica y Cosmolog\'ia'', IFCA-CSIC / Universitat de Val\`encia, Spain\\
$^{15}$ Observat\'orio Nacional/MCTI, Rua General Jos\'e Cristino, 77, São Crist\'ov\~ao, CEP 20921-400, Rio de Janeiro, Brazil\\
$^{16}$ Centro de Estudios de F\'isica del Cosmos de Arag\'on (CEFCA), Unidad Asociada al CSIC, Plaza San Juan 1, 44001, Teruel, Spain\\
$^{17}$ Department of Astronomy, University of Michigan, 311 West Hall, 1085 South University Ave., Ann Arbor, USA\\
$^{18}$ University of Alabama, Department of Physics and Astronomy, Gallalee Hall, Tuscaloosa, AL 35401, USA\\
$^{19}$ Universidade de S\~ao Paulo, Instituto de Astronomia, Geof\'isica e Ci\^encias Atmosf\'ericas, Depto. de Astronomia, Rua do Mat\~ao, 1226, CEP 05508-090, S\~ao Paulo, Brazil\\
$^{20}$ Instruments4, 4121 Pembury Place, La Canada Flintridge, CA 91011, USA
}


\appendix

\section{Magnitude shifts} \label{appx:magr_shift}

\begin{figure}
\includegraphics[width=\linewidth]{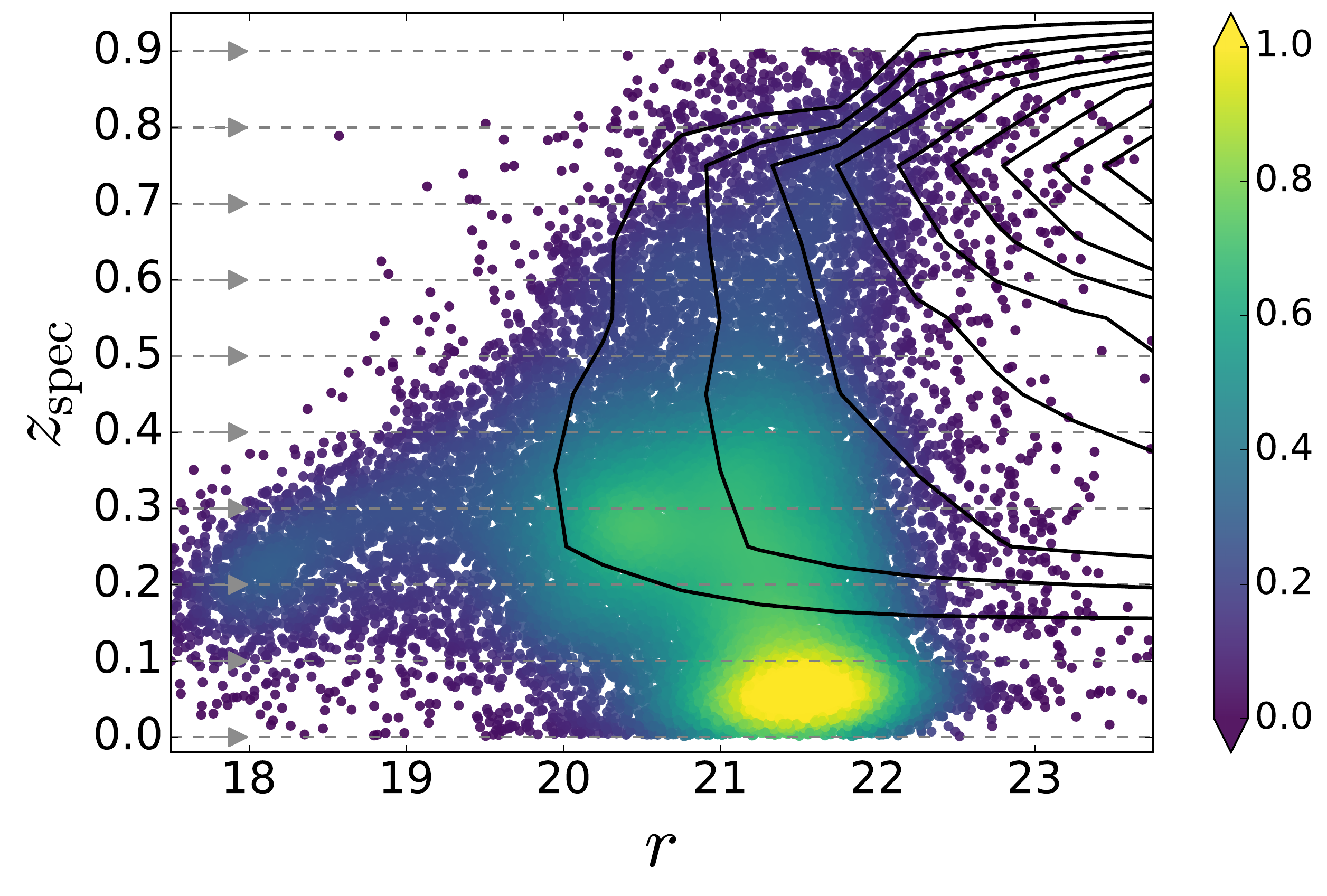}
\caption{Magnitude-redshift distributions for the galaxies from the Superset catalogue (colour map) and the galaxy luminosity function (contour lines). The arrows on the left side (at fixed redshift values) indicate the direction in which the magnitude shifts occur. The colour code represents the normalized density of galaxy spectra.}
\label{fig:magr_z_gal}
\end{figure}

In Fig. \ref{fig:magr_z_gal} we illustrate the procedure of shifting the original $r$-band magnitudes of the SDSS DR12Q galaxy spectra in order to meet the targeted number densities per deg$^{2}$ required by the luminosity function down to the faint end, while preserving the spectroscopic identification (i.e. redshift) of the sources. For a fixed value in the y-axis, we have randomly generated several realizations of the same spectrum at fainter magnitudes so as to reach the same number densities as expected from the GLF (in a given simulated area).

We are limited by the number of SDSS spectra available in some bins, as clearly illustrated here. In particular, we lack a sufficiently representative sample of galaxy spectra at $z>0.3$ for all magnitudes. This implies that, in the different sets of mocks, we may found several ``duplicates'', i.e. pseudo spectra that were generated from the same SDSS spectrum but using different magnitude seeds. Although being more critical for galaxies, this effect is noticeable for all types of sources. None the less, we have verified that the presence of these so-called duplicates in the mocks does not bias the performance of the classifiers.

\section{Comparison of observations and simulated fluxes for different noise models} \label{appx:hist_delta_flam}

The same analysis shown in Fig. \ref{fig:hist_delta_flam_6bands} is provided in Figs. \ref{fig:hist_delta_flam_1} and \ref{fig:hist_delta_flam_2} for the remaining 54 bands.

\begin{figure*}
\includegraphics[width=\textwidth]{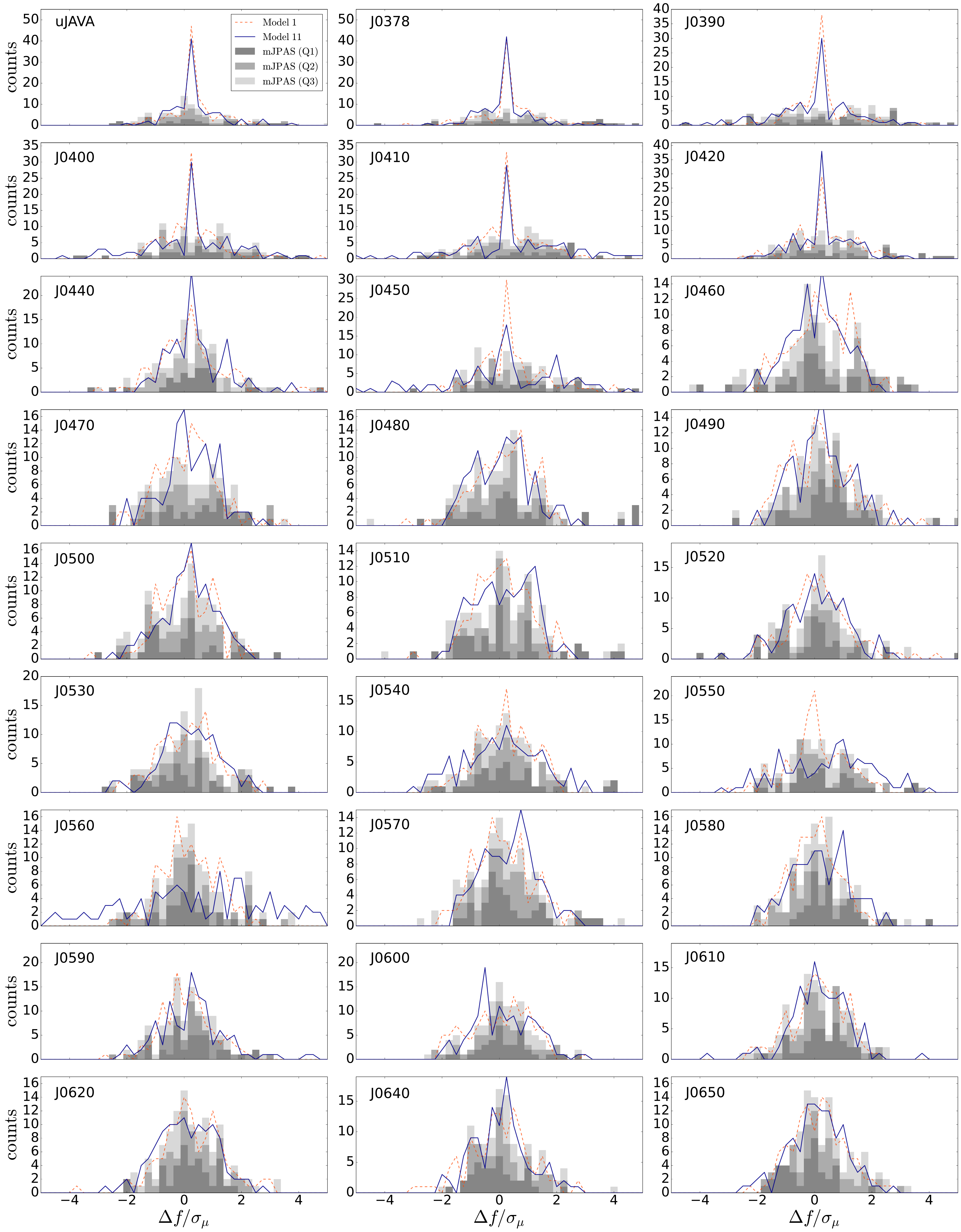}
\caption{Histograms of the differences ($\Delta f$) between the observed (or simulated) and the synthetic fluxes divided by the nominal errors ($\sigma_{\mu}$) for the spectroscopic stars. The bars correspond to the miniJPAS observations, which were divided into lower (Q1), median (Q2) and upper (Q3) tertiles, shown by the different shades of gray ranging from darker to lighter. We compare two different noise models: 1 (orange dashed lines) and 11 (blue solid lines).}
\label{fig:hist_delta_flam_1}
\end{figure*}

\begin{figure*}
\includegraphics[width=\textwidth]{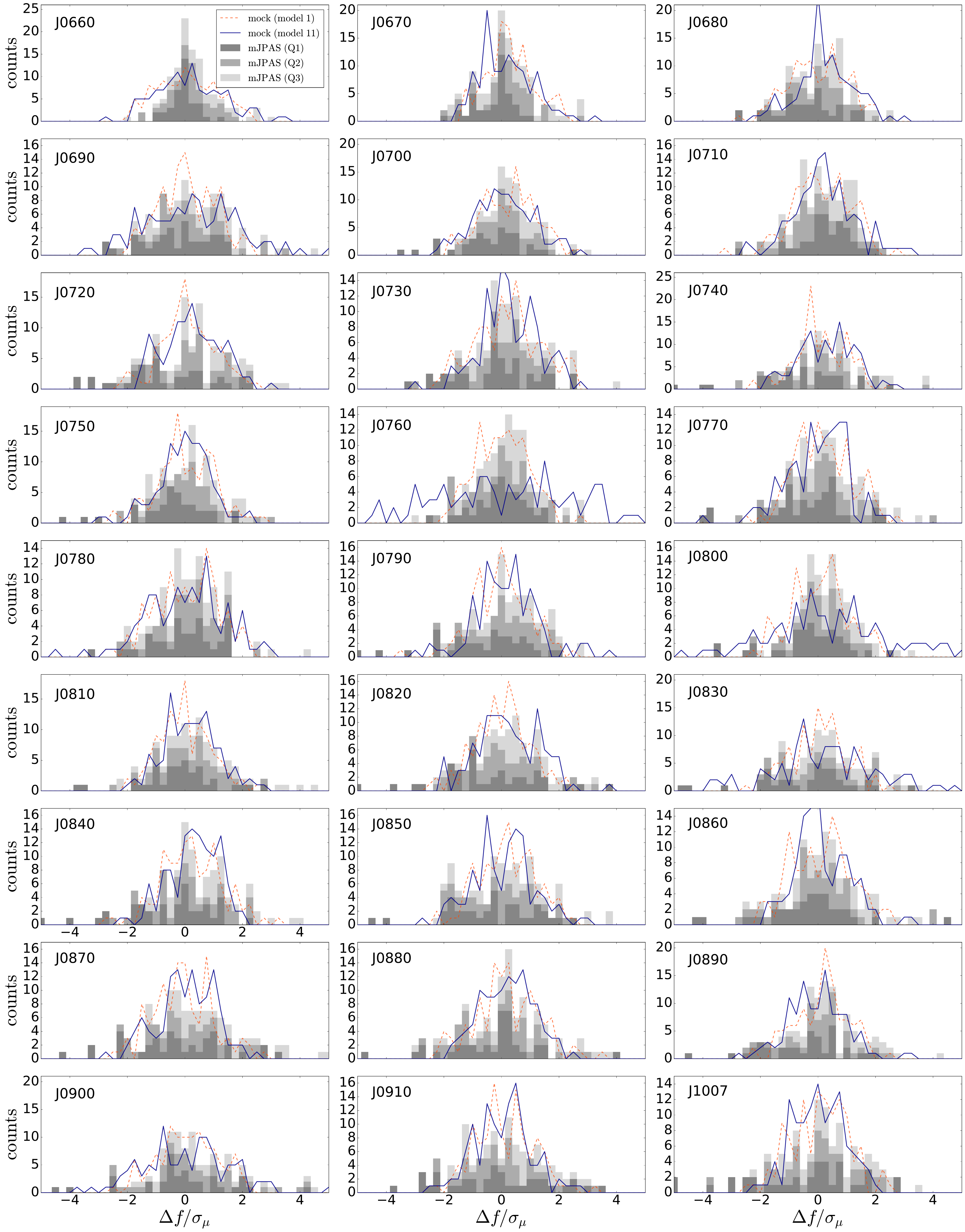}
\caption{continued.}
\label{fig:hist_delta_flam_2}
\end{figure*}
\bsp	
\label{lastpage}
\end{document}